\documentclass[groupedaddress,showpacs,nofootinbib]{revtex4}
\usepackage{graphicx}
\usepackage{bm}
\usepackage{amssymb}
\usepackage{amsmath}
\usepackage{epsfig}
\usepackage{hhline}
\usepackage{color}
\usepackage{multirow}
\usepackage[normalem]{ulem}

\def\lapp{\mathrel{\rlap{\raise.5ex\hbox{$<$}}
                    {\lower.5ex\hbox{$\sim$}}}}
\def\gapp{\mathrel{\rlap{\raise.5ex\hbox{$>$}}
                    {\lower.5ex\hbox{$\sim$}}}}

{\newcommand{\lsim}{\mbox{\raisebox{-.6ex}{~$\stackrel{<}{\sim}$~}}}
{\newcommand{\gsim}{\mbox{\raisebox{-.6ex}{~$\stackrel{>}{\sim}$~}}}

\newcommand{\bmt}{\begin{pmatrix}}
\newcommand{\emt}{\end{pmatrix}}
\newcommand{\ba}{\begin{array}{c}}
\newcommand{\ea}{\end{array}}
\newcommand{\be}{\begin{equation}}
\newcommand{\ee}{\end{equation}}
\newcommand{\bea}{\begin{eqnarray}}
\newcommand{\eea}{\end{eqnarray}}

\newcommand{\bi}{\begin{itemize}}
\newcommand{\ei}{\end{itemize}}

\newcommand{\baz}{\begin{array}{cc}}

\newcommand{\mathsym}[1]{{}}

\newcommand{\bt}{\begin{tabular}}
\newcommand{\et}{\end{tabular}}

\newcommand{\benu}{\begin{enumerate}}
\newcommand{\eenu}{\end{enumerate}}
\newcommand{\bav}{\begin{array}{cccc}}

\begin{document}
\title{\bf Minimal Vector-like leptonic Dark Matter and Signatures at the LHC} 
\author{Subhaditya Bhattacharya}
\email{subhab@iitg.ernet.in}
\affiliation{Department of Physics, Indian Institute of Technology Guwahati, North Guwahati, Assam- 781039, India}
\author{Nirakar Sahoo}
\email{nirakar.pintu.sahoo@gmail.com}
\author{Narendra Sahu}
\email{nsahu@iith.ac.in}
\affiliation{Department of Physics, Indian Institute of Technology, Hyderabad, Yeddumailaram, 502205, 
Telengana, India}

\begin{abstract}
We propose a minimal vector-like leptonic dark matter (DM) with renormalisable interaction in a beyond the Standard Model (SM) 
scenario, where the SM is augmented with a vector-like doublet and a singlet leptons. The additional fermions are odd under a 
discrete $Z_2$ symmetry, while the rest of the SM particles are singlets, and thus providing stability to the DM. In this scenario, 
we show that, the DM emerges as an admixture of the neutral component of the vector-like doublet and the singlet leptons. The singlet-doublet 
mixing ($\sin \theta$) plays a crucial role in yielding the correct relic density as well as in obtaining null direct DM search results 
through an interplay of interactions via $Z$ and Higgs mediation. The mixing is also strongly constrained from the invisible Z and Higgs 
decay width. We found that the correct relic abundance of DM can be obtained in a large region of parameter space for DM-mass larger than 
$ M_Z/2$ and $\sin \theta \lsim 0.1$. The details of model phenomenology with collider signatures at the Large hadron Collider (LHC) are 
discussed. In particular, we show that for $\sin \theta \lsim 0.01$, the charged companion of the DM can give rise to an observable 
displaced vertex signature, marking a significant departure from other fermionic DM scenarios, while keeping the relic abundance intact.   
\end{abstract}

\pacs{98.80.Cq,12.60.Jv}
\maketitle

\newpage

\section{Introduction}
Convincible evidences from galaxy rotation curve, gravitational lensing and large scale structures imply that there exist dark matter (DM) 
in the present Universe~\cite{DM_review1,DM_review2}. These evidences imply only the gravitational interaction of DM and hence its direct 
detection has remained a mystery yet. However, the relic abundance of DM is precisely measured by WMAP~\cite{wmap} and recently by 
PLANCK~\cite{PLANCK}. The question is what kind of particle constitutes DM ? The Standard Model (SM) does not include any particle candidate 
of DM. This hints towards new physics beyond the SM (BSM). 

The existence of large scale structures of the Universe implies that the DM should be either stable or its life time is longer than 
the age of the Universe. Moreover, the observed abundance of DM points out that the freeze-out cross-section of DM to be around 
$\langle \sigma|v|\rangle \approx 3\times 10^{-26}{\rm cm}^3/{\rm s \equiv 1 pb}$, which is a typical weak interaction cross-section. 
Therefore, it is usually assumed that a stable weakly interacting massive particle (WIMP) of mass in the sub-TeV region is a suitable 
candidate of DM~\cite{kolb&turner}. In the early Universe, the DM is assumed to be in thermal equilibrium via its weak interaction 
cross-sections. As the temperature falls below the mass scale of DM, the latter gets decoupled from the thermal bath. Since then the DM 
density, measured in terms of $n_{\rm DM}/s$, where $s$ is the entropy density, remains constant and is given by: 
\begin{equation}
Y_{\rm DM}^{\rm obs}\equiv \frac{n_{\rm DM}}{s} \approx 4\times 10^{-12} \left( \frac{100 {\rm GeV}}{M_{\rm DM}} \right) 
\left( \frac{\Omega_{\rm DM} h^2}{0.1196}\right)\,.
\end{equation} 
Apart from the relic abundance, we don't have any clue about the DM properties, such as its mass, spin {\it etc}. This leads to a large number 
of models in the BSM scenario which can correctly produce the relic abundance of DM. However, these models are strongly constrained by the 
null observation of DM in direct detection at terrestrial experiments such as Xenon-100~\cite{xenon100} and LUX~\cite{lux}. In particular, the 
current upper limit on the spin-independent WIMP-nucleon elastic cross-section set by LUX~\cite{lux} is given to be $7.6\times 10^{-46} {\rm cm}^2$ 
at a WIMP mass of 33 GeV. 

A vector-like colourless fermion with zero hypercharge is a simple possibility to be considered as a candidate of DM in the 
BSM scenario. These fermions are similar to SM leptons even though they may not carry any leptonic charge. The singlet and triplet 
leptons with hypercharge (Y) zero need an extra symmetry for their stability, while a quintet fermion with $Y=0$ is stable by 
itself~\cite{Cirelli:2005uq}. The neutral component of these fermions can be a viable candidate of DM. On the other hand, the neutral component of 
a vector-like doublet or a quartet lepton with non-zero hypercharge can not qualify itself to be a candidate of DM even in presence of an 
extra symmetry due to its large Z-mediated WIMP-nucleon elastic cross-section. However, a vector-like doublet DM can be reinstated in presence 
of a heavy scalar triplet~\cite{nsahu_papers}, where the relic abundance mostly arises from an asymmetric component while annihilating the 
symmetric component efficiently to the SM particles. 

In the above said scenarios,  a singlet lepton ($\chi^0$), odd under a $Z_2$ symmetry, is a minimal possibility 
for a candidate of DM. However, without introducing additional fields, such DM can only have an effective interaction to the SM via a 
dimension five operator of the form $\bar{\chi^0}\chi^0 H^{\dagger}H/\Lambda$, where $\Lambda$ is unknown. In order to obtain a full 
theory of such an operator, one needs to introduce additional scalar singlets or vector-like lepton doublets. In the former case, 
a scalar singlet, say $\eta$, mixes with the SM Higgs ($h$) and thus yielding an effective Yukawa interactions of the fermionic DM to 
SM through terms like $\bar{\chi^0}\chi^0 \eta \to \bar{\chi^0}\chi^0 h$ (see for example \cite{Kim}). In the latter case, the additional 
vector-like lepton doublets are assumed to be odd under the same $Z_2$ symmetry as the singlets. As a result the DM emerge as an admixture 
of the neutral component of the doublets and the singlet~\cite{ArkaniHamed:2005yv,Mahbubani:2005pt,D'Eramo:2007ga,Enberg:2007rp,Cynolter:2008ea,Cohen:2011ec,
Cheung:2013dua,Restrepo:2015ura,Calibbi:2015nha,Cynolter:2015sua,F-DM-recent}, which we consider below.     

In this paper we study in details the possibility of a minimal fermion DM with renormalisable interaction by adding a vector-like lepton 
doublet $N^T=(N^0,N^-)$ and a singlet $\chi^0$ to the SM particle spectrum in such a way that the DM emerge as an admixture of $N^0$ and 
$\chi^0$. The stability of the DM is ensured by imposing an extra $Z_2$ symmetry under which both the new fermions are odd, while all other 
SM particles are even. Interactions of the fermionic DM with SM is mainly dictated by a Higgs Yukawa and 
through $Z$ interactions of the doublet component. We demonstrate that the singlet-doublet mixing plays an important role in constraining 
the parameter space of the DM model in obtaining correct relic density and null direct search result through an interesting interplay of 
the $Z$ and Higgs mediated interactions.  The strongest constraint on the singlet-doublet mixing comes from the invisible Z-decay width measured 
at LEP for $M_{DM} < M_Z/2$. Moreover, the direct search of DM at Xenon-100 and LUX restricts the mixing angle: $\sin \theta \lsim 0.1$ for its 
mass above 45 GeV. However, these small values of $\sin \theta$ do not affect the relic abundance of DM. We show that observed DM abundance can 
be obtained for its mass larger than 45 GeV and even $\sin \theta \lsim 0.1$. In most of the parameter space, coannihilation plays a dominant 
role in yielding the current relic density of DM.  We notice that, the small values of $\sin \theta$ give an interesting signature of the charged 
companion of the DM, {\it i.e.} $N^\pm$. For example, the three body decay of $N^\pm$ can give observable displaced vertex signature at the Large 
Hadron Collider (LHC) if the mass splitting between DM and its charged partner is less than 10 GeV and $\sin \theta < 0.01$.   

The advantage of the model considered here is its minimal number of parameters ({\it i.e.} three) and very rich phenomenology. The model 
has sufficient parameter space left to be explored in direct DM search and collider experiments, while it can be utilised to explain the 
recently observed 750 GeV di-photon excess at LHC~\cite{Bhattacharya:2016lyg} or can be connected to neutrino sector to explain non-zero 
$\theta_{13}$~\cite{Bhattacharya:2016lts} with minimum addition of fields. Another advantage of considering vector-like fermions as DM is 
that they do not introduce any extra anomaly to the SM. The additional vector-like fermions have been largely studied in the literature in 
different contexts~\cite{vectorlike_fermion,vectorlike_fermion_DM,vectorlike_fermion_collider}. The properties of these new fermions are subject 
to the tight constraints from electroweak precision measurements and by direct searches~\cite{Cynolter:2008ea,EW_constraint1,EW_constraint2}. 

The paper is arranged as follows. In section-II, we discuss the important aspects of the DM model. In section-III, we estimate the constraints on 
model parameters from the invisible Z and Higgs decay. Section-IV is devoted to find constraints on model parameters from the spin independent 
DM-nucleon cross-section. Section-V is devoted to estimate the relic density of DM. In section-VI we calculate the constraints on model parameter 
from electroweak precision tests. Section-VII is devoted to lay some predictions of the model at LHC. We conclude in section-VIII.

\section{The Model, parameters and interactions}
Let us extend the SM with two vector-like fermions: a doublet $N^T(\equiv (N^0, N^-))$ (1,2,-1) and a singlet $\chi^0$ (1,1,0), 
where the numbers inside the parentheses are the quantum numbers corresponding to the SM gauge group $SU(3)_c\times SU(2)_L\times U(1)_Y$. In 
addition to that we impose a discrete symmetry $Z_2$ under which $N$ and $\chi^0$ are odd, while all other fields are even \footnote{For a 
generic discussion on DM stability versus global symmetry, see \cite{DM-stability}.}. As a result the DM emerge as an admixture of $N^0$ 
and $\chi^0$. The relevant Lagrangian can be given as:
\begin{equation}\label{eq:Lag-lepton}
\hspace*{-0.5cm}
 -\mathcal{L_{\rm Yuk}}  \supset  M_N \overline{N}N + M_\chi \overline{\chi^0}\chi^0 + \left[ Y\overline{N}\widetilde{H}\chi^0 
+ {\rm h.c.}\right]\,,
\end{equation}
where $M_N$ and $M_{\chi}$ are mass parameters corresponding to the doublet and singlet vector like leptons and $Y$ denotes the interaction strength 
among them. Note here that due to vector-like nature, mass terms are perfectly gauge invariant. In Eq. (\ref{eq:Lag-lepton}), $\widetilde{H}=i\tau_2 H^*$, 
where $H$ is the SM Higgs iso-doublet: $H = \begin{pmatrix} H^+ \cr \\
H_0 \,
\end{pmatrix}$. After electroweak phase transition, the vacuum expectation value (vev) of SM Higgs: $\langle H \rangle= \begin{pmatrix} 0 \cr \\
v \,
\end{pmatrix}$ gives rise to a mixing between $N^0$ and $\chi^0$. In the basis $(\chi^0, N^0)$, the mass matrix is given by
\begin{equation}
\mathcal{M} = \begin{pmatrix} M_\chi & m_D\cr \\
m_D & M_N 
\end{pmatrix}\,.
\end{equation} 
where $m_D=Y v$ and $v=174$ GeV. Diagonalizing the above mass matrix we get two mass eigenvalues:
\begin{eqnarray}
M_1 \approx M_\chi-\frac{m_D^2}{M_N-M_\chi}\nonumber\\
M_2 \approx M_N + \frac{m_D^2}{M_N-M_\chi}\label{mass-eigenstates}
\end{eqnarray}
where we have assumed $m_D << M_N, M_\chi$. The corresponding mass eigenstates are given by:
\begin{eqnarray}
N_1= \cos \theta \chi^0 + \sin \theta N^0\nonumber\\
N_2=\cos \theta N^0 - \sin \theta \chi^0\,,
\end{eqnarray}
where the mixing angle is:
\begin{equation}\label{mixing}
\tan 2\theta = \frac{2 m_D}{M_N-M_\chi}\,. 
\end{equation}
Note that $N_2$ is dominantly a doublet with a small admixture of singlet component. On the other hand, $N_1$ is dominantly a 
singlet with a small admixture of doublet component, which makes it a viable candidate of DM.

\begin{figure}[htbp]
	\includegraphics[width=.60\textwidth]{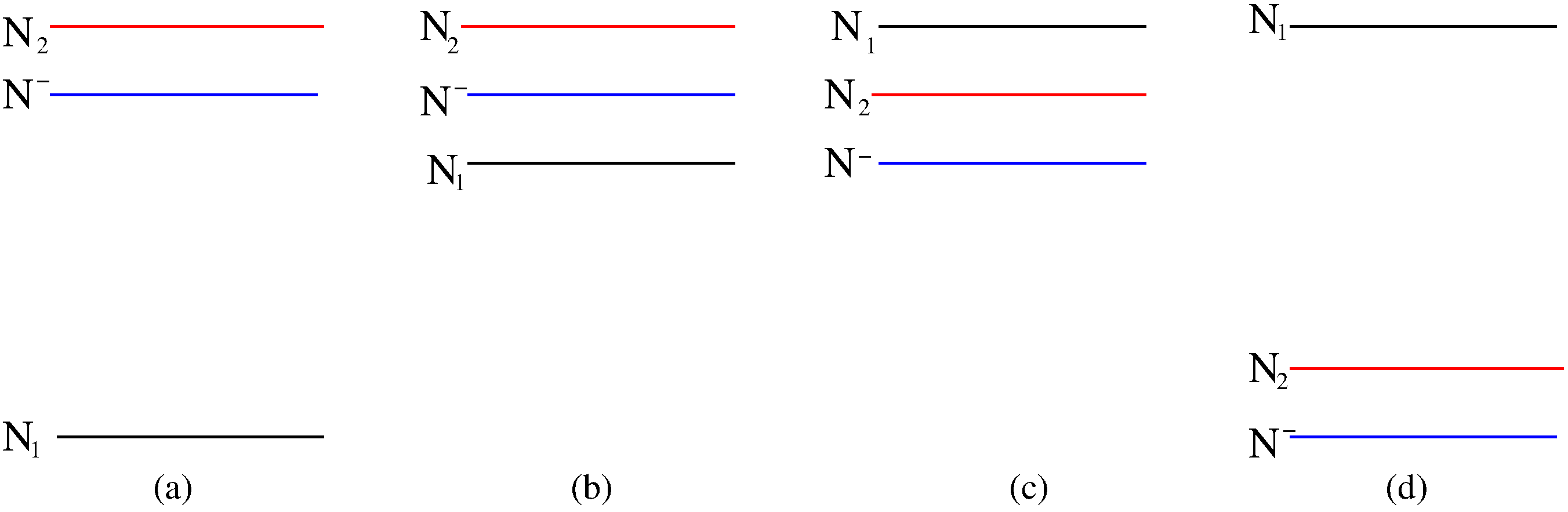}
	\caption{Pictorial presentation of the possible mass spectrum for additional vector-like leptons.}\label{mass_spectrum}
\end{figure}

In the physical spectrum we also have a charged vector-like fermion $N^\pm$ whose mass in terms of $M_1$ and $M_2$ and the mixing 
angle $\theta$ can be given as:
\begin{equation}\label{chargedfermion_mass}
M^\pm =M_1 \sin^2 \theta + M_2 \cos^2 \theta \simeq M_N\,.
\end{equation}
We will see later that the allowed values of the mixing angle is quite small, {\it i.e.} $\sin \theta \lsim 0.1$. Therefore, we have $M_2 \approx M_N$. 
This implies that the vector-like lepton $N^\pm$ is almost degenerate to neutral vector-like lepton $N_2$, $M_2 \approx M^\pm$. Since $\sin \theta \sim 0.1$, 
we always get $M_N < M_2$ unless $M_1$ is quite large, say $M_1 > {\cal O} (10^4)$ GeV. For $M_1 \lsim  {\cal O}(10^4)$ GeV and $\sin \theta \lsim 0.1$, 
we can have four possibilities in the mass spectrum of additional vector-like leptons as shown in the Fig. \ref{mass_spectrum}. From Fig. \ref{mass_spectrum} 
(c) and (d) we see that the charged lepton is the lightest stable fermion and hence excluded from DM consideration. So, the remaining possibilities are 
Fig. \ref{mass_spectrum} (a) and (b), where $N_1$ is the lightest stable particle (LSP) and is a suitable DM candidate. The next to lightest stable 
particle (NLSP) is the charged vector-like fermion $N^-$ and next to next lightest stable particle (NNLSP) is $N_2$. The mass splitting between $N_1$ and 
$N^-$ is $(M_N-M_\chi)+m_D^2/(M_N-M_\chi)$, where as the mass splitting between $N_1$ and $N_2$ is $(M_N-M_\chi)+2 m_D^2/(M_N-M_\chi)$. Depending on the choice of 
$M_1$ and $M_2$, the mass splitting between $N_1$ and $N_2$ can be either large (Fig. \ref{mass_spectrum} (a)) or small (Fig.\ref{mass_spectrum} (b)).

Let us now turn to the interaction terms in the mass basis of $N_1$ and $N_2$. The Yukawa interaction term can be re-written as: 
\begin{eqnarray}
Y\overline{N} \widetilde{H} \chi^0 + {\rm h.c.} & \to &  Y \overline{N^0} h \chi^0 + {\rm h.c.}\nonumber\\
&=&  Y\left[ \sin 2\theta(\overline{N_1} h N_1 - \overline{N_2} h N_2 )+ \cos 2\theta(\overline{N_1} h N_2 + \overline{N_2} h N_1) \right]\,.
\end{eqnarray}
Similarly the charge current and neutral current gauge interaction in the mass basis of $N_1$ and $N_2$ can be given as:
\begin{equation}
\frac{g}{\sqrt{2}} \overline{N^0} \gamma^\mu W_\mu^+ N^- + {\rm h.c.} \to \frac{g\sin \theta}{\sqrt{2}} 
\overline{N_1}\gamma^\mu W_\mu^+ N^- + \frac{g\cos \theta}{\sqrt{2}} \overline{N_2}\gamma^\mu W_\mu^+ N^- + {\rm h.c.}\,,
\end{equation} 
\begin{equation}
\frac{g}{2\cos \theta_w} \overline{N^0}\gamma^\mu Z_\mu N^0  \to \frac{g}{2\cos \theta_w} \left( \sin^2 \theta \overline{N_1} \gamma^\mu Z_\mu N_1 
+ \sin \theta \cos \theta ( \overline{N_1} \gamma^\mu Z_\mu N_2 + \overline{N_2} \gamma^\mu Z_\mu N_1) + \cos^2 \theta \overline{N_2} \gamma^\mu Z_\mu N_2 \right) \,.  
\end{equation}
The neutral current of $N^-$ is not affected by the singlet-doublet mixing and is given by:
\begin{equation}
e\gamma^\mu \overline{N^-} A_\mu N^- + \frac{g}{2}\overline{N^-} \gamma^\mu Z_\mu N^-\,.
\end{equation}

Essentially, the model contains three independent parameters in terms of 
\bea
\{M_1, M_2, \sin \theta \,\} ~~~{\rm or}~~~\{Y, M_1, M_2 \}
\eea
where, $Y$ and $\sin \theta$ are related by  
\bea
Y= \frac{\Delta M sin 2\theta}{2v}\,,
\eea
as seen from Eq. (\ref{mixing}). In the following sections, we will derive limits on these parameters from different 
experimental and theoretical constraints. Then we will discuss the possible signatures which can be used as a probe to our model.

\section{Constraint on model parameters from invisible decays}

\subsection{ Invisible Z-decay}
\begin{figure}[htbp]
\includegraphics[scale=0.4]{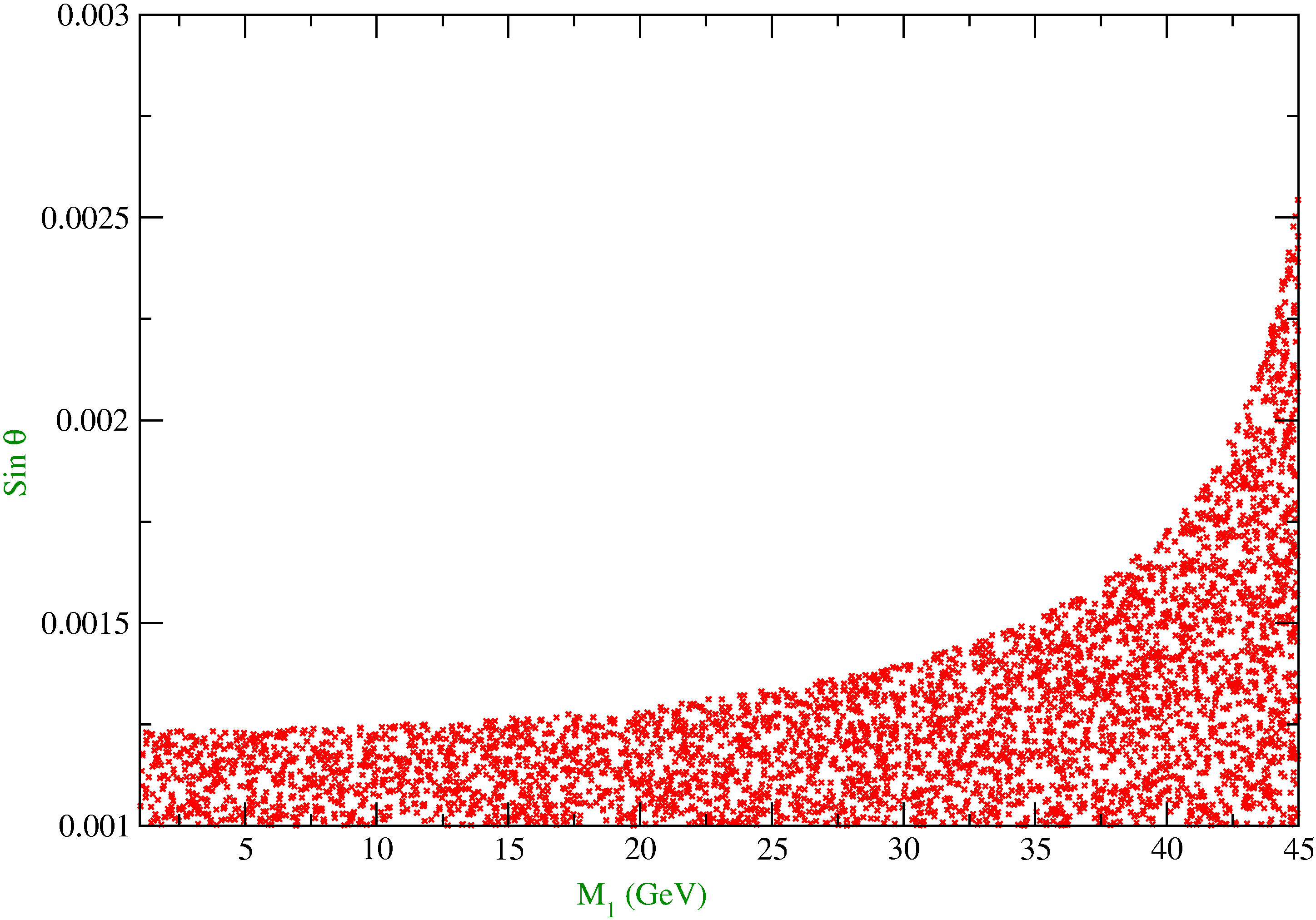}
\caption{The allowed values of $\sin \theta$ for different dark matter mass $M_1 < M_Z/2=45$ GeV from invisible $Z$ decay constraints. We assume here 
$M_2=M_z/2=$ 45 GeV.}\label{z-decay}
\end{figure}

The non observation of $Z$ decay width to a fourth generation charged lepton pairs prohibit to $M^\pm > M_z/2$. As $M^\pm \simeq M_2 \simeq M_N$, 
this implies that the mass of $N^-$ and $N_2$ has to be larger than 45 GeV. On the other hand $M_1$ can be as light as 1 GeV~\cite{lee_weinberg_bound}. 
Due to singlet-doublet mixing, the Z-boson can decay to $N_1$ and $N_2$. Since $N_2$ is heavier than $M_Z/2$, the decay of $Z$ to $N_2 N_2$ 
is also forbidden. Hence the relevant decay widths of the processes $Z\to N_1 N_1$ and $Z\to N_1 N_2$ can be given as: 
\begin{eqnarray} 
\Gamma (Z\to N_1 N_1) &=& \frac{1}{48\pi }M_Z  \left( \frac{ g^2 \sin^4 \theta}{\cos^2 \theta_w}\right) \left( 1 + \frac{2 M_1^2}{M_Z^2} \right) 
\left(1-\frac{4 M_1^2}{M_Z^2} \right)^{1/2}\nonumber\\
\Gamma (Z\to N_1 N_2) &=& \frac{1}{96\pi} M_Z \left( \frac{ g^2 \sin^2 \theta \cos^2 \theta}{\cos^2 \theta_w}\right) \left( \left( 1- \frac{ (M_1^2+M_2^2)}{M_Z^2}\right) + \frac{6 M_1 M_2}{M_Z^2} + \left( 1- \frac{(M_1^2 -M_2^2)^2}{M_Z^4} \right) \right) \nonumber \\
&&\left(1-2\frac{(M_1^2+M_2^2)}{M_Z^2}+\frac{(M_1^2 - M_2^2)^2}{M_Z^4} \right)^{1/2}\,
\end{eqnarray} 
The invisible Z-decay width in the standard model is $\Gamma({\rm invisible})= 499 \pm 1.5 $MeV~\cite{pdg}. Therefore, if $Z$ is allowed to decay to 
$N_1 N_1$ and $N_1 N_2$ then the decay width should not be more than 1.5 MeV. Under this condition we have shown the constraints on $\sin \theta$ for 
various values of $M_1$ in Fig. (\ref{z-decay}), while fixing $M_2=M_z/2=$ 45 GeV, the minimum possible value, for simplicity. We see that the DM mass 
$M_1$ can be allowed below $M_z/2=$ 45 GeV only if $\sin \theta < 10^{-3}$. 

\subsection{Invisible Higgs decay}
\begin{figure}[htbp]
\includegraphics[scale=.4]{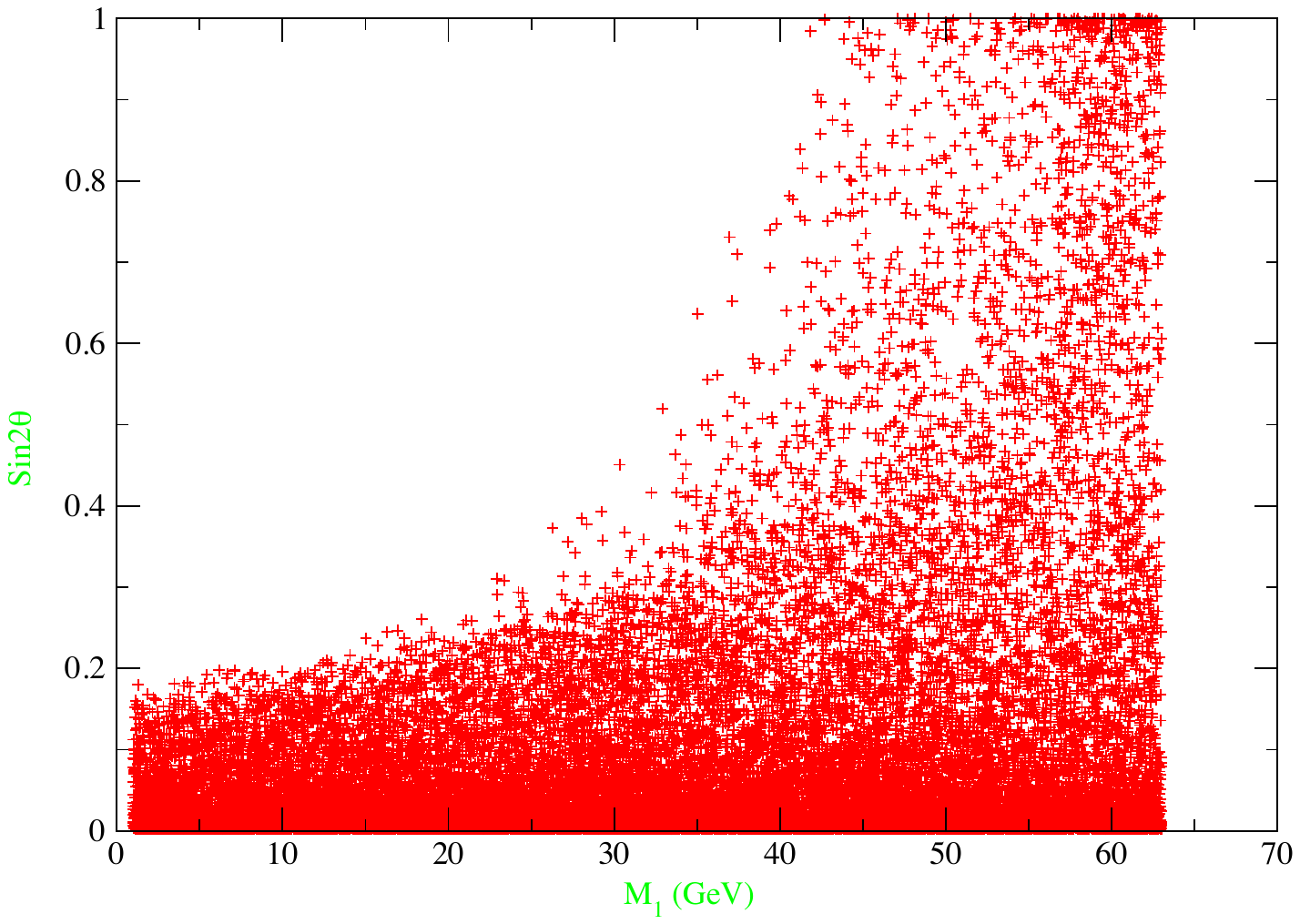}
\caption{Scatter plot for allowed parameter space in $\sin 2\theta-M_1$ (GeV) plane from invisible Higgs decay. All possible values of $M_2$ used that keeps the decay chain open.}\label{higgs_decay}
\end{figure}

The SM Higgs can decay to $N_1$ and $N_2$ and therefore strongly constrained by the observation. In particular, 
the branching ratio for the invisible Higgs decay width is given by
\begin{equation}
{\rm Br}_{\rm inv} = \frac{\Gamma_h^{\rm inv}}{\Gamma_h^{\rm SM} +\Gamma_h^{\rm inv} }  \,,
\end{equation}
where $\Gamma_h^{\rm SM}=4 $MeV. The invisible Higgs decay width is given by:
\begin{equation}
\Gamma_h^{\rm inv}=\Gamma(h \to N_1 N_1) + \Gamma(h \to N_2 N_2) +\Gamma(h \to N_1 N_2) 
\end{equation} 
where 
\begin{eqnarray}
\Gamma(h \to N_i N_i) &=& \frac{(Y\sin 2\theta)^2}{8\pi} M_h \left(1-\frac{4M_i^2}{M_h^2} \right)^{3/2}\nonumber\\
\Gamma(h \to N_i N_j) &=& \frac{(Y\cos 2\theta)^2}{8\pi} M_h \left( 1 - \frac{M_i^2 +M_j^2}{M_h^2} -\frac{2 M_i M_j}{M_h^2} \right)\nonumber\\
 &\times & \left(1-\frac{2(M_i^2 +M_j^2)}{M_h^2}+ \frac{(M_i^2-M_j^2)^2}{M_h^4} \right)^{1/2}\,.
\end{eqnarray}
Taking ${\rm Br}_{\rm inv} < 0.3$~\cite{invisible_higgs} we have shown the allowed region in the plane of $\sin 2\theta$ versus $M_1$ in 
Fig. (\ref{higgs_decay}). We saw that for small DM mass, typically $1 < M_1 < M_H/2=63$ GeV, $\sin 2\theta$ is strongly constrained, while 
for $M_1 \gsim  63$ GeV, large $\sin 2\theta$ is allowed from invisible Higgs decay constraints. In the scan, we are choosing all possible 
values of $M_2$ that keeps the decay chain open. Clearly, the invisible $Z-$ decay puts stronger constraint on the mixing angle than the 
invisible decay of Higgs does.

\section{Implication on model parameters from direct search of dark matter}
\begin{figure}[thb]
$$
\includegraphics[height=5.5cm]{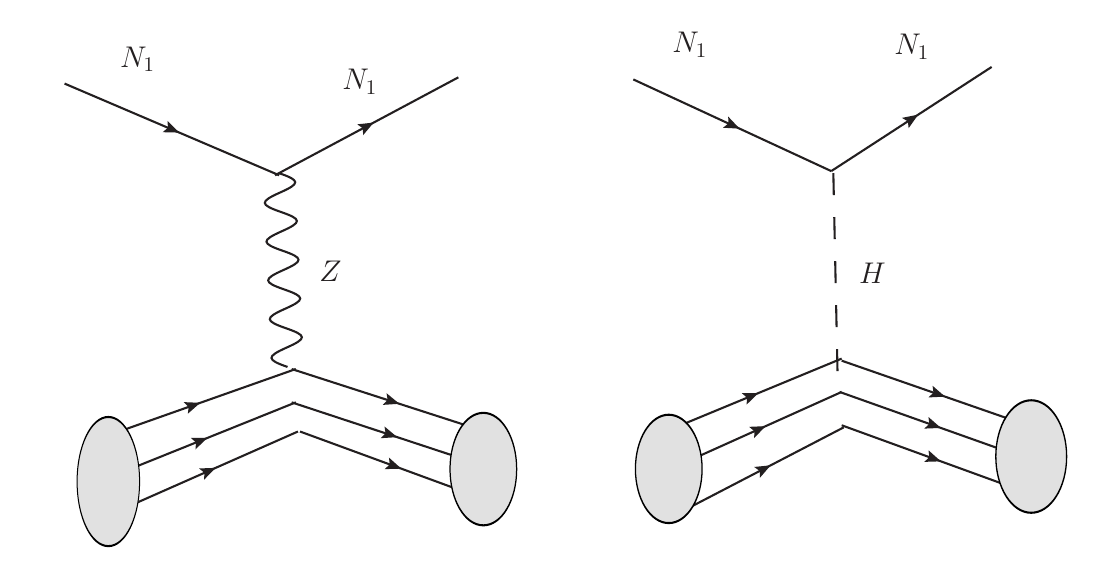}
$$
\caption{Feynman diagrams for direct detection of $N_1$ DM.}
\label{fig:DD2}
\end{figure}

\begin{figure}[htbp]
	\includegraphics[width=.40\textwidth]{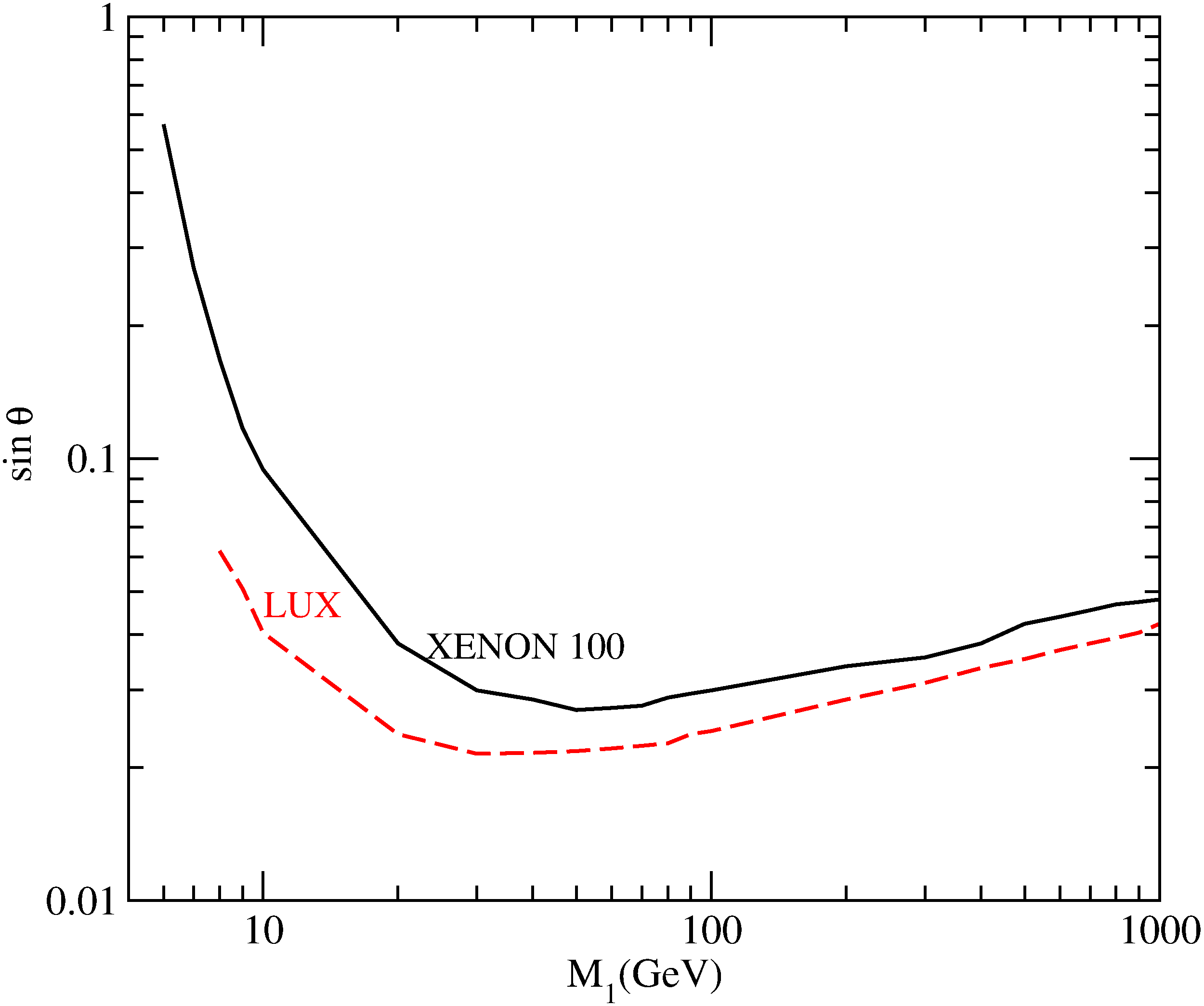}
        \includegraphics[width=.40\textwidth]{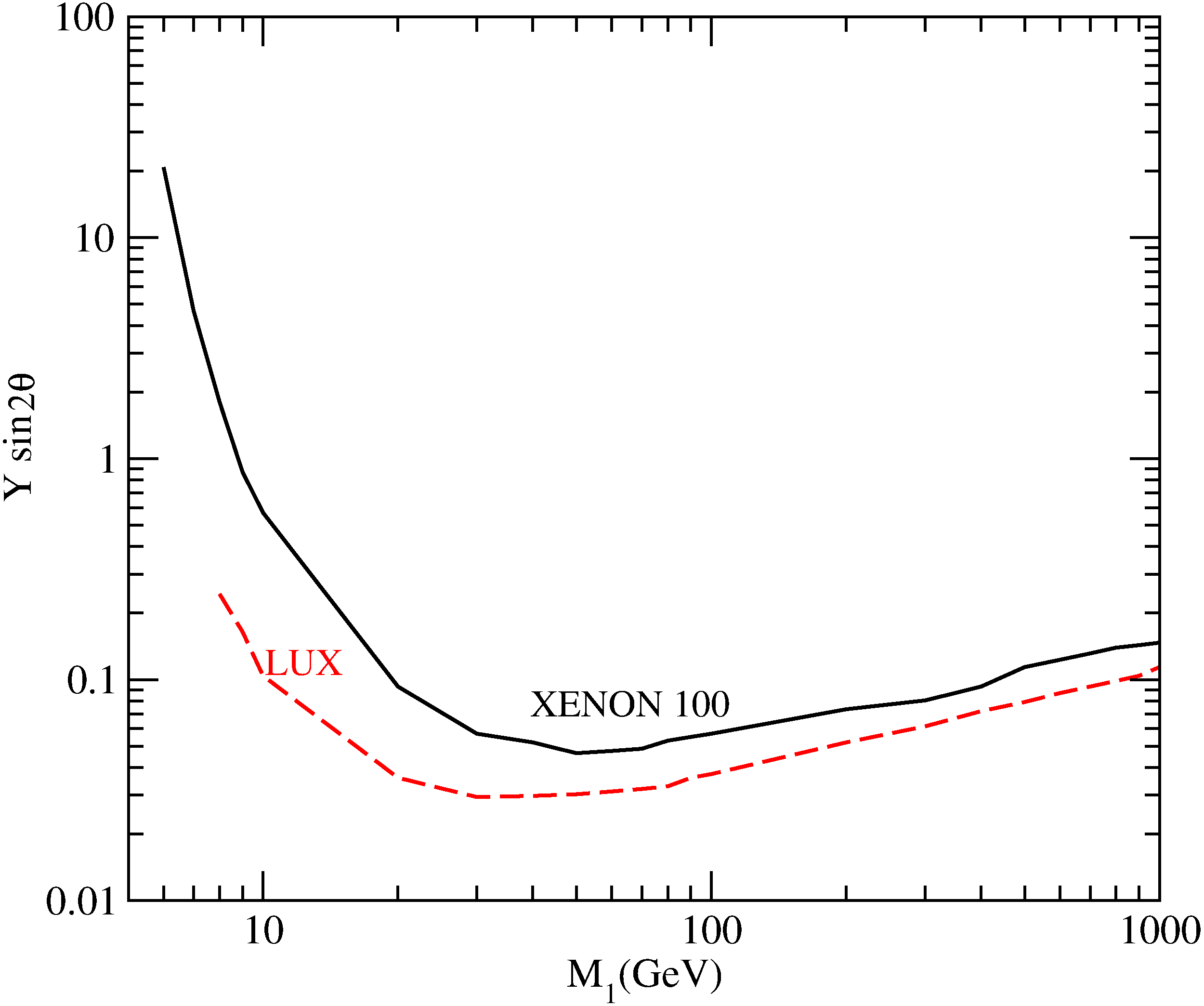}
	\caption{Constraint on $\sin \theta$ (left) from $Z$ mediated direct detection process and $Y\sin 2\theta$ (right) from $H$ mediated direct 
         detection process using Xenon-100 and LUX data for different values of DM mass $M_1$ . }\label{sintheta_constraint}
\end{figure}

We shall now point out constraints on the model parameters from direct search of DM. The relevant diagrams through which $N_1$ interacts with the nuclei are 
shown in Fig. (\ref{fig:DD2}). In particular, our focus will be on Xenon-100~\cite{xenon100} and LUX~\cite{lux} which at present give strongest constraint 
on spin-independent DM-nucleon cross-section having seen no DM event yet. In our model, this in turn puts a stringent constraint on the mixing angle 
$\sin \theta$ for spin independent DM-nucleon interaction mediated via the $Z$-boson (see in the left of Fig. (\ref{fig:DD2})). The cross-section per 
nucleon for $Z$ mediation is given by~\cite{ Goodman:1984dc,Essig:2007az}
\begin{equation}\label{DM-nucleon-Z}
\sigma_{\rm SI}^Z = \frac{1}{\pi A^2 }\mu_r^2 |\mathcal{M}|^2   
\end{equation}
where $A$ is the mass number of the target nucleus, $\mu_r=M_1 m_n/(M_1 + m_n) \approx m_n$ is the reduced mass, $m_n$ is the mass of nucleon 
(proton or neutron) and $\mathcal{M}$ is the amplitude for $Z$-mediated DM-nucleon cross-section given by
\begin{equation}\label{Z-mediated-process}
\mathcal{M}= \sqrt{2} G_F [Z (f_p/f_n) + (A-Z)] f_n \sin^2 \theta\,,
\end{equation}
where $f_p$ and $f_n$ are the interaction strengths (including hadronic uncertainties) of DM with proton and neutron respectively 
and $Z$ is the atomic number of the target nucleus. For simplicity we assume conservation of isospin, {\it i.e.} $f_p/f_n=1$. The value of 
$f_n$ vary within a range: $0.14 < f_n < 0.66$~\cite{hadronic_uncertainty}. If we take $f_n \simeq 1/3$, the central value, then from 
Eqs. (\ref{DM-nucleon-Z}) and (\ref{Z-mediated-process}), we get the Z-mediated cross-section per nucleon to be 
\begin{equation}
\sigma_{\rm SI}^Z \simeq 3.75 \times 10^{-39} {\rm cm}^2 \sin^4 \theta\,.
\end{equation}
In the above equation the only unknown is the $\sin \theta$ and hence can be constrained from observation. Using the data from Xenon-100 and LUX 
we have shown the allowed values of $\sin \theta$ in the left panel of Fig. (\ref{sintheta_constraint}) as a function of the DM mass. 

Another possibility of having spin-independent DM-nucleon interaction is through the exchange of SM Higgs (see in the right of Fig. (\ref{fig:DD2})). 
The cross-section per nucleon is given by:
\begin{equation}\label{scalar_mediated_crossssection}
\sigma_{\rm SI}^h=\frac{1}{\pi A^2}\mu_r^2  \left[ Z f_p + (A-Z)f_n \right]^2
\end{equation} 
where the effective interaction strengths of DM with proton and neutron are given by:
\begin{equation}\label{f-values}
f_{p,n} = \sum_{q=u,d,s}f_{Tq}^{(p.n)} \alpha_q \frac{m_{(p,n)}}{m_q} + \frac{2}{27} f_{TG}^{(p,n)}\sum_{q=c,t,b} \alpha_q \frac{m_{p.n}}{m_q}
\end{equation}
with 
\begin{equation}\label{alpha-value}
\alpha_q = \frac{ Y\sin 2\theta}{M_h^2} \left( \frac{m_q}{v}\right) \,.
\end{equation}
In Eq. (\ref{f-values}), the different coupling strengths between DM and light quarks are given by~\cite{DM_review1} $f^{(p)}_{Tu}=0.020\pm 0.004$, 
$f^{(p)}_{Td}=0.026\pm 0.005$,$f^{(p)}_{Ts}=0.118\pm 0.062$, $f^{(n)}_{Tu}=0.014\pm 0.004$,$f^{(n)}_{Td}=0.036\pm 0.008$,$f^{(n)}_{Ts}=0.118\pm 0.062$. 
The coupling of DM with the gluons in target nuclei is parameterized by 
\begin{equation}\label{Gluon-interaction}
f^{(p,n)}_{TG}=1-\sum_{q=u,,d,s}f^{(p,n)}_{Tq}\,. 
\end{equation}
Thus from Eqs. (\ref{scalar_mediated_crossssection},\ref{f-values},\ref{alpha-value},\ref{Gluon-interaction}) the spin-independent 
DM-nucleon cross-section is given to be:
\begin{equation}
\sigma_{\rm SI}^h=\frac{4}{\pi A^2}\mu_r^2 \frac{Y^2 \sin^2 2\theta}{M_h^4} 
\left[\frac{m_p}{v}\left( f_{Tu}^p + f_{Td}^p + f_{Ts}^p+ \frac{2}{9}f_{TG}^p\right) + \frac{m_n}{v} \left(f_{Tu}^n + f_{Td}^n + 
f_{Ts}^n+ \frac{2}{9}f_{TG}^n \right)\right]^2\,.
\end{equation}
In the above equation the only unknown quantity is $Y$ or $\sin 2\theta$ which can be constrained by requiring that $\sigma_{\rm SI}^h$ is 
less than the current DM-nucleon cross-sections at Xenon-100 and LUX. This is shown in the right panel of Fig. (\ref{sintheta_constraint}).
\begin{figure}[thb]
$$
\includegraphics[height=5.5cm]{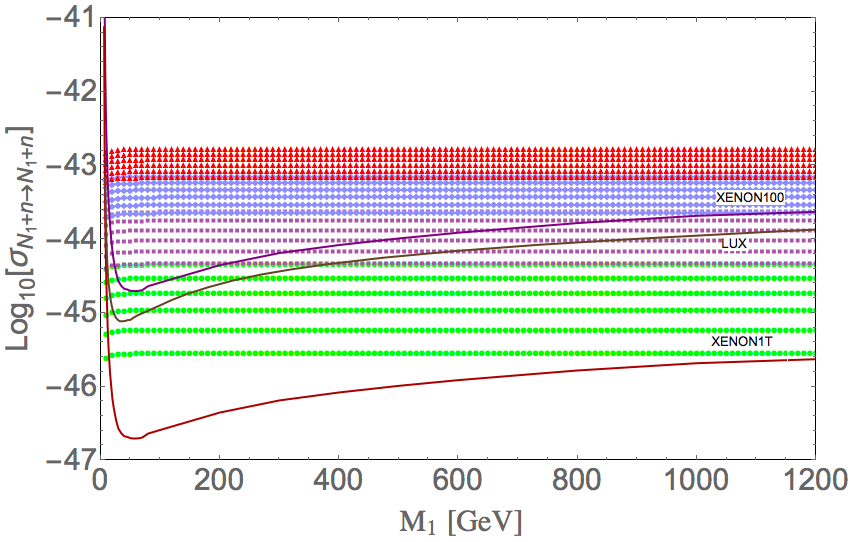}
\includegraphics[height=5.5cm]{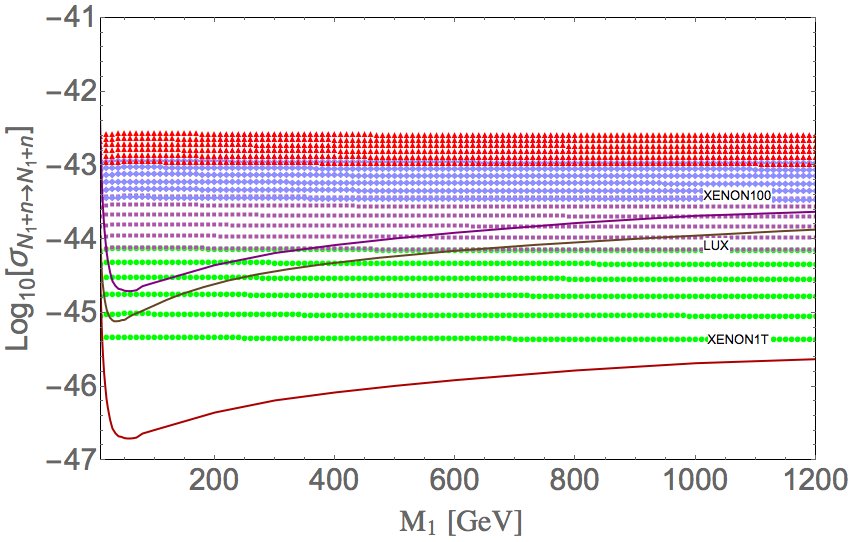}
$$
\caption{Spin independent direct detection cross-section for $N_1$ DM as a function of DM mass for $\sin \theta=\{0.05-0.1\}$ (Green), $\sin \theta=\{0.1-0.15\}$ (Purple), $\sin \theta=\{0.15-0.2\}$ (Lilac), $\sin \theta=\{0.2-0.25\}$ (Red). Fixed values of $\Delta M =\{100,700\}$ GeV (left and right respectively) have been used. XENON 100, LUX data are shown with XENON 1 T prediction.}
\label{fig:DD1}
\end{figure}

Now we make a combined analysis by taking both $Z$ and $H$ mediated diagrams taken into account together. We vary mixing angle $\sin \theta$, as well as the mass of DM with fixed $\Delta M$ and hence changing $Y=\Delta M \sin 2\theta/2v$ accordingly. In Fig \ref{fig:DD1}, we show the spin-independent cross-section for $N_1$ DM within its mass range $M_{1}: 50-1200$ GeV. The plot is obtained by varying 
$\sin \theta$ within $\{0.05-0.25\}$ with $\sin \theta=\{0.05-0.1\}$ (Green), $\sin \theta=\{0.1-0.15\}$ (Purple), $\sin \theta=\{0.15-0.2\}$ (Lilac), $\sin \theta=\{0.2-0.25\}$ (Red) by choosing a fixed set of $\Delta M =\{100,700\}$ GeV (left and right 
respectively). It clearly shows that the larger is $\sin \theta$, the stronger is the interaction strength (through larger contribution from $Z$ mediation) and hence the larger is the DM-nucleon cross-section. Similarly, the larger is $\Delta M$, the greater is the $Y$-value and hence larger is the DM-nucleon cross-section (through larger contribution from Higgs mediation). Hence, it turns out that direct search experiments constraints $\sin \theta$ to a great extent. For example, we see that with $\sin \theta=0.1$, the DM mass $M_1> 350$ GeV. The effect of $\Delta M$ on DM-nucleon 
cross-section is much smaller as we can see from the left and right panel of Fig. (\ref{fig:DD1}). However, we note that $\Delta M$ plays a dominant role in the relic abundance of DM. Approximately, $\sin \theta \le 0.1$ (Green points) are allowed for most of the parameter space except for smaller DM masses. Further small mixing angles are still allowed and that will have a non-trivial outcome in collider search.

We would also like to note here that there is a very tiny amount of spin-dependent cross-section arises through $Z$ mediation, but the cross-section lies far far below than the observed limit and hence it effectively doesn't constrain the parameter space at all. For example, with $\{M_1=80 {\rm GeV},M_2=120 {\rm GeV},\sin\theta=0.1\}$, the spin dependent cross-section for proton is as low as $3.2\times 10^{-49}$ pb compared to $4.2\times 10^{-9}$ pb for spin independent one.

\section{Relic abundance of $N_1$ dark matter}
\begin{figure}[thb]
$$
\includegraphics[height=3.0cm]{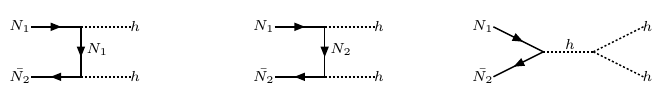}
$$
$$
\includegraphics[height=2.7cm]{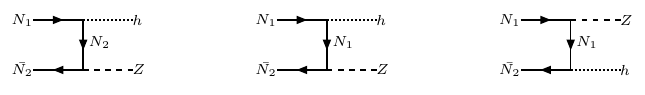}
$$
$$
\includegraphics[height=3.0cm]{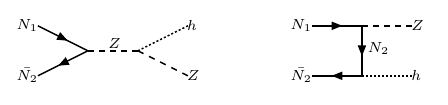}
$$
$$
\includegraphics[height=2.8cm]{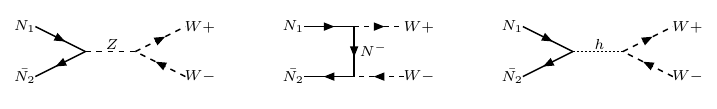}
$$
$$
\includegraphics[height=2.7cm]{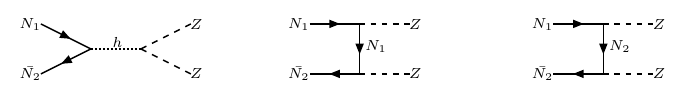}
$$
$$
\includegraphics[height=3.0cm]{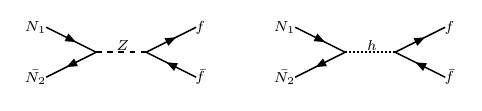}
$$

\caption{Dominant Coannihilation processes with $N_2$ to Higgs, gauge boson pair and $f\bar{f}$, where $f$ stands for all fermions.}
\label{fd-coann1}
\end{figure}

\begin{figure}[thb]
$$
\includegraphics[height=1.7cm]{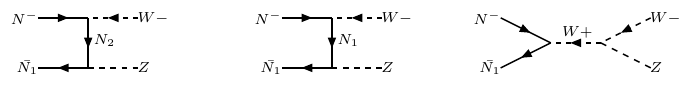}
\includegraphics[height=1.7cm]{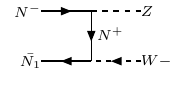}
$$
$$
\includegraphics[height=2.7cm]{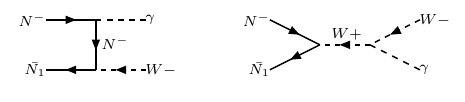}
$$
$$
\includegraphics[height=3.0cm]{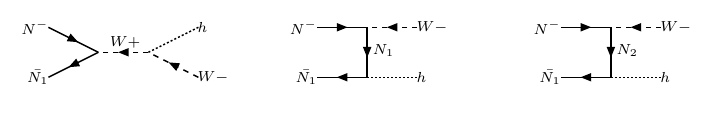}
$$
$$
\includegraphics[height=3.0cm]{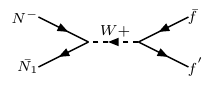}
$$
\caption{Dominant co-annihilation contributions from $N_{1}N^{-}$ to Gauge boson and Higgs productions along with $f^{'}\bar{f}$, where $f$ stands for all the SM fermions.}
\label{fd-coann2}
\end{figure}

\begin{figure}[thb]
$$
\includegraphics[height=3.0cm]{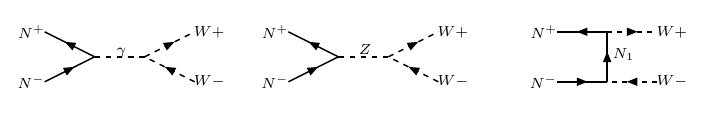}
$$
$$
\includegraphics[height=3.1cm]{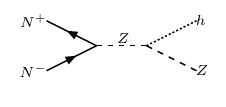}
\includegraphics[height=3.1cm]{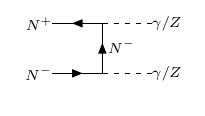}
$$
$$
\includegraphics[height=3.1cm]{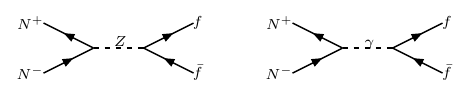}
$$
\caption{Dominant co-annihilation contributions from $N^{+}N^{-}$ to Gauge boson and Higgs productions along with $f\bar{f}$, where $f$ stands for all the SM fermions.}
\label{fd-coann3}
\end{figure}

In order to estimate the relic abundance of the $N_1$ DM, we need to calculate the various cross-sections through which $N_1$ 
abundance depletes. 
\begin{figure}[thb]
$$
\includegraphics[height=3.3cm]{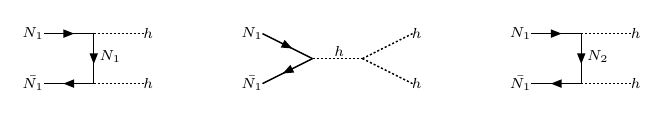}
$$
$$
\includegraphics[height=3.1cm]{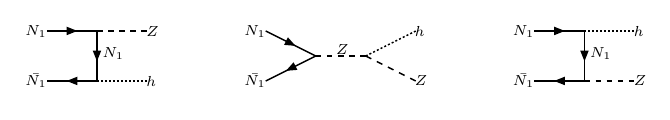}
$$
$$
\includegraphics[height=3.3cm]{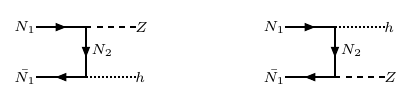}
$$
$$
\includegraphics[height=2.6cm]{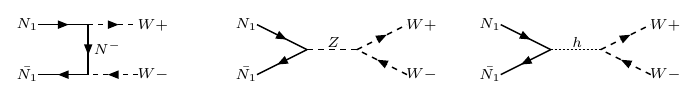}
$$
$$
\includegraphics[height=2.6cm]{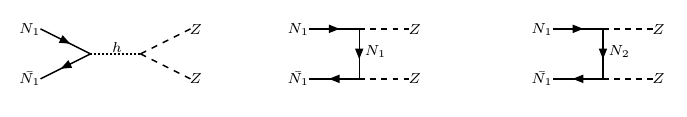}
$$
$$
\includegraphics[height=2.9cm]{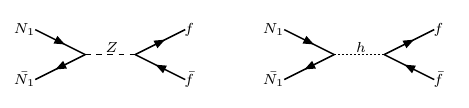}
$$

\caption{Dominant Annihilation processes to Higgs and gauge boson productions along with $f\bar{f}$, where $f$ stands for all the SM fermions.}
\label{fd-ann}
\end{figure}
The main annihilation processes have been indicated in Fig. \ref{fd-ann}. The dominant channels are $\overline{N_1} N_1 \to h h$ and $\overline{N_1} N_1 \to W^+ W^-$. As direct detection of DM restricts us to have a mixing angle: $\sin\theta \lsim 0.1$, the annihilations channel: $\overline{N_1} N_1 \to W^+ W^-$ can give large cross-section. The other relevant channels are mainly coannihilation of $N_1$ with 
$N_2$ and $N^\pm$. We have shown $\overline{N_1} N_2 \to SM$ in Fig \ref{fd-coann1}. The annihilation of $\overline{N_2} N_2 \to SM$ is very 
similar to $\overline{N_1} N_1 \to SM$ and are not shown explicitly. If $N_1$ is degenerate to $N^{\pm}$, then we find co annihilations of $\overline{N_1} N^{\pm} \to SM$ (in Fig. \ref{fd-coann2}), $\overline{N_2} N^{\pm} \to SM$ (similar to $\overline{N_1} N^{\pm} \to SM$) and $N^{\mp}N^{\pm} \to SM$ (in Fig. \ref{fd-coann3}) are also important for correct relic density of DM.

The relic density of the $N_1$ DM can be given by~\cite{griest}
\begin{equation}
\Omega_{N_1} h^2 = \frac{1.09 \times 10^9 \rm GeV^{-1}}{g_\star ^{1/2} M_{PL}} \frac{1}{J(x_f)}\,,
\end{equation}
where $J(x_f)$ is given by
\begin{equation}
J(x_f)= \int_{x_f}^ \infty \frac{\langle \sigma |v| \rangle _{eff}}{x^2} \hspace{.2cm} dx \,,
\end{equation}
where $\langle \sigma |v| \rangle _{eff}$ is the thermal average of dark matter annihilation cross sections including contributions from coannihilations as follows:
\begin{equation}\label{sigma_eff}
\begin{split}
\langle \sigma |v| \rangle _{eff} = & \frac{g_1^2}{g_{eff}^2} \sigma (\overline{N_1} N_1)
 +2 \frac{g_1 g_2}{g_{eff}^2} \sigma (\overline{N_1} N_2) (1+\Delta)^{3/2} exp(-x\Delta) \\
 & +2 \frac{g_1 g_3}{g_{eff}^2} \sigma (\overline{N_1} N^-) (1+\Delta)^{3/2} exp(-x\Delta)\\
 &+2 \frac{g_2 g_3}{g_{eff}^2} \sigma (\overline{N_2} N^-) (1+\Delta)^{3} exp(-2x\Delta)
 + \frac{g_2 g_2}{g_{eff}^2} \sigma (\overline{N_2} N_2) (1+\Delta)^{3} exp(-2x\Delta) \\
& + \frac{g_3 g_3}{g_{eff}^2} \sigma (N^+ N^-) (1+\Delta)^{3} exp(-2x\Delta)\,.
\end{split}
\end{equation}
In the above equation $g_1$,$g_2$ and $g_3$ are the spin degrees of freedom for $N_1$, $N_2$ and $N^-$ respectively. Since these are spin half 
particles, all g's are 2. The freeze-out epoch of $N_1$ is parameterized by $x_f= \frac{M_{1}}{T_f}$, where $T_f$ is the freeze out temperature. 
$\Delta$ depicts the mass splitting ratio as $\Delta = \frac{M_{i}- M_{1}}{ M_{1}} $, where $M_i$ stands for the mass of $N_2$ and $N^{\pm}$. The 
effective degrees of freedom $g_{eff}$ in Eq. (\ref{sigma_eff}) is given by
 \begin{equation}
 g_{eff} = g1+ g_2 (1+\Delta)^{3/2} exp(-x\Delta) + g_3 (1+\Delta)^{3/2} exp(-x\Delta)\,.
\end{equation}  

Note that the dark-sector, spanned by the $Z_2$ odd vector-like fermions, is mainly dictated by three independent 
parameters: 
\begin{equation}\label{parameter-space}
\sin\theta, M_{1}, M_{2}\,.
\end{equation}
For small $\sin \theta$ the mass splitting between $N^-$ and $N_2$ is almost less than a GeV. Therefore, for all practical purpose 
$N_2$ and $N^-$ masses can be taken to be same. However, in the numerical calculation we have determined $M^\pm$ using 
Eq. (\ref{chargedfermion_mass}). In the following we vary the parameters given in Eq. (\ref{parameter-space}) and find the allowed 
region of correct relic abundance for $N_1$ DM satisfying WMAP~\cite{wmap} constraint ~\footnote{The range we use corresponds to the 
WMAP results; the PLANCK constraints $0.112 \leq \Omega_{\rm DM} h^2 \leq 0.128$~\cite{PLANCK}, though more stringent, do not lead to 
significant changes in the allowed regions of parameter space.}
\bea
0.094 \leq \Omega_{\rm DM} h^2 \leq 0.130 \,.
\label{eq:wmap.region}
\eea
\begin{figure}[thb]
$$
\includegraphics[height=5.5cm]{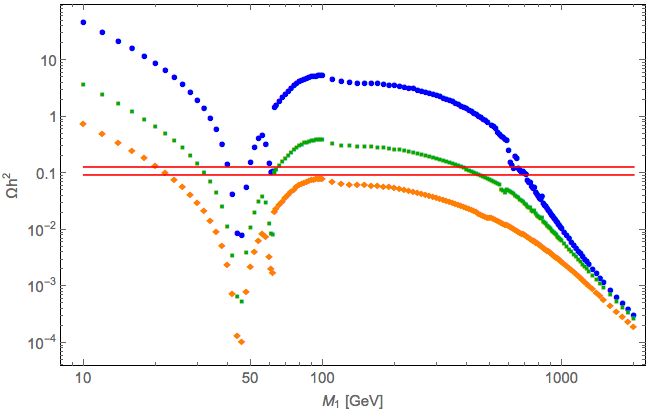}
\includegraphics[height=5.5cm]{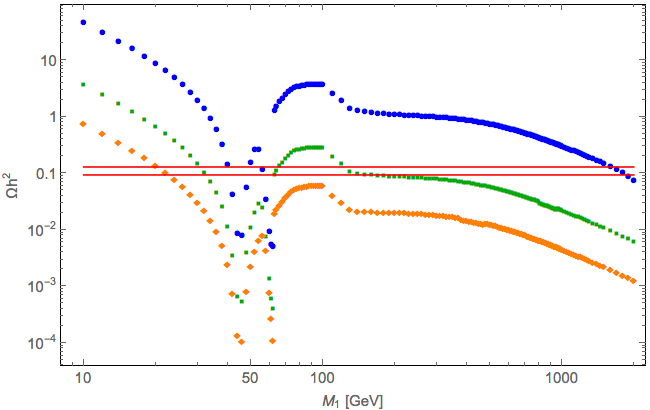}
$$
\caption{Variation of relic density with  DM mass ( $M_1$ in GeV) for $\sin\theta =\{0.1,0.2,0.3\}$ (Blue, Green and Orange respectively from top to bottom). 
On the left panel we have taken $M_{2}-M_{1}=$ 100 GeV while on the right panel we have set $M_{2}-M_{1}=$ 500 GeV. Red band indicates relic density within 
WMAP range as given in Eq. \ref{eq:wmap.region}.}
\label{fig:Omega-m-1}
\end{figure}

\begin{figure}[thb]
$$
\includegraphics[height=6.0cm]{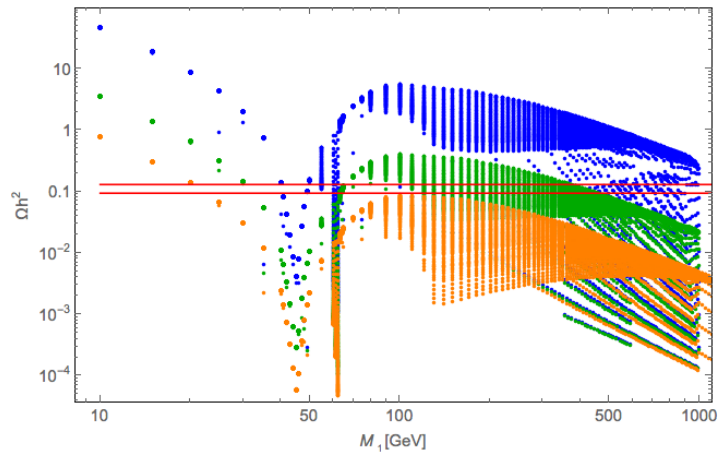}
$$
\caption{ Same as Fig. \ref{fig:Omega-m-1}, but with $\Delta M=\{10-1000\}$ GeV, $\sin\theta =\{0.1,0.2,0.3\}$ is indicated in Blue, Green and Orange respectively. Red band indicates relic density within WMAP range as given in Eq \ref{eq:wmap.region}.}
\label{fig:Omega-m-2}
\end{figure}

The parameter space scan presented here has been generated using the code MicrOmegas~\cite{micro}. In Fig. \ref{fig:Omega-m-1}, we show how relic 
density changes with DM mass for different choices of mixing angle: $\sin \theta=\{0.1 (\rm Blue), 0.2 (\rm Green), 0.3 (\rm Orange)\}$, with fixed 
mass difference $\Delta M = M_{2}-M_{1}=$ 100 GeV (left panel) and 500 GeV (right panel). First of all, we see that for a fixed $\sin \theta$ the relic density decreases with increasing DM masses. This means the annihilation and co-annihilation together increases with increasing DM masses. This feature is mainly due to dominant contribution to annihilation through $Z$ to $W^{+}W^-$ final state which increases with increasing DM mass. Secondly, we see that as the mixing angle increases, the relic abundance decreases. This is because the cross-sections mediated by $Z$ boson (say for example, $\overline{N_1} N_1 \to W^+ W^-$) increases due to larger $SU(2)$ component with increasing $\sin \theta$. Note that invisible $Z$-decay width restricts $\sin \theta < 0.001$ for $M_1 < 45 $ GeV. However, in this region we always have $\langle \sigma|v|\rangle_{\rm eff} < \langle \sigma|v|\rangle_{\rm freeze-out}$, leading to large DM abundance. For $M_1 > 45$ GeV, large $\sin \theta$ values are allowed from invisible $Z$ decays and hence the scans presented here are not in contrast with that. 
From Fig. \ref{fig:Omega-m-1} we also see that in the low mass region around $M_{1}\simeq 45$ GeV, there is a sharp drop due to the resonant annihilation of $N_1$ to SM particles through $Z$. The Higgs resonance is also clearly seen at a DM mass $M_1 \simeq M_h/2= 63$ GeV. Naturally, the right hand side of Fig. \ref{fig:Omega-m-1} with larger $\Delta M$ and hence a larger Yukawa $Y$ shows more prominent Higgs resonance.  As the mass splitting: $\Delta M$ increases, the relic density shifts to higher value for a given $\sin \theta$ due to smaller annihilation cross-section due to $N_2$ propagation as well as smaller co-annihilation contribution. These features all together control the allowed range of parameter space for correct relic density of DM. In Fig. \ref{fig:Omega-m-2}, we show the whole parameter space for relic density as a function of $M_1$. In this case $\Delta M$ has been varied between $10-1000$ GeV with fixed $\sin \theta=\{0.1 (\rm Blue), 0.2 (\rm Green), 0.3 (\rm Orange)\}$. Basically, Fig. \ref{fig:Omega-m-1} refers to a subset of the parameter space of Fig. \ref{fig:Omega-m-2}, which indicates that $\sin \theta \gsim 0.3$ is almost ruled out by relic abundance constraint. The reason is that with larger singlet-doublet mixing, the annihilation cross-section is dominated by $\overline{N_1} N_1 \to W^+ W^-$ and hence we always get a smaller relic density than the desired value.

\begin{figure}[htbp]
$$
\includegraphics[height=5.5cm]{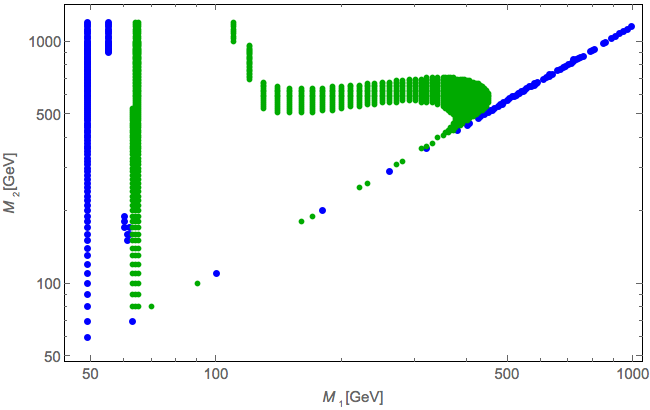}
\includegraphics[height=5.5cm]{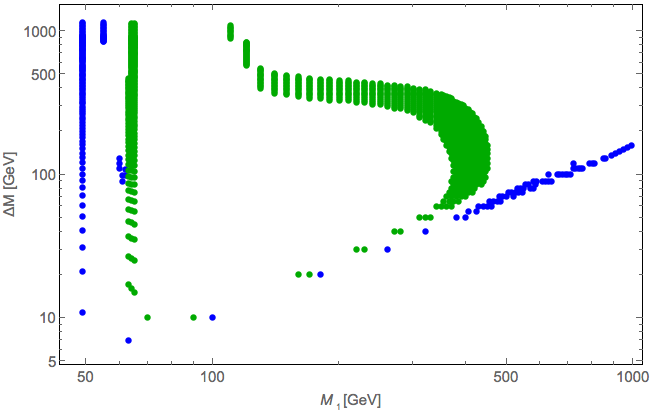}
$$
\caption{ Left: Dark matter mass $M_1$ versus $M_2$ (all in GeV) to satisfy relic density constraint. The allowed region is shown in blue and green for $\sin\theta=\{0.1,0.2\}$ respectively. Right: Same as left but in terms of $\Delta M = M_2-M_1$ vs $M_{1}$. }
\label{fig:m1-m2}
\end{figure}

Thus from Fig. (\ref{fig:Omega-m-1}) and (\ref{fig:Omega-m-2}), we show that $\sin \theta \gsim 0.3$ can not satisfy relic density constraint. Therefore, in what follow we restrict ourselves to $\sin \theta=0.1,0.2$ for the remaining analysis. In left panel of 
Fig. \ref{fig:m1-m2}, we show a scattered plot in the plane of DM mass $M_{1}$ and $M_2$ for $\sin\theta=\{0.1, 0.2\}$ (Blue and Green respectively) that yields the observed DM abundance. On the right panel, the same plot is recast in the plane of $\Delta M = M_2-M_1$ versus $M_1$. From Fig. \ref{fig:m1-m2}, we see that for small values of mixing, say $\sin \theta=0.1$, $N_1$ is dominantly a singlet and  the annihilation channels, as shown in Fig. \ref{fd-ann}, heavily depends on the singlet-doublet mixing. Hence for small mixing, say $\sin\theta=0.1$, the annihilations produce smaller cross-section than the required to satisfy relic abundance. Hence, 
in this case, correct relic density is obtained only if co-annihilations compensate for the rest of required cross-sections. This is clearly shown by Blue line which lives close to the boundary $M_2\simeq M_1$, yielding larger coannihilations and hence correct DM relic density. On the other hand, for a relatively larger mixing, say $\sin\theta=0.2$ as shown in Green dots in Fig. \ref{fig:m1-m2}, there are essentially two 
different regions that satisfy relic density for a given DM mass. For small $M_1$, the region with small $\Delta M$ has a moderate annihilation cross-section with smaller Yukawa coupling ($Y \propto \Delta M $), while that is compensated by appropriate amount of coannihilation. On the other hand, larger $\Delta M$ also satisfy relic density with larger Yukawa coupling making the annihilation cross-section itself in the right ball park. With increasing DM mass the annihilation cross-section increases due to unitarity behaviour of $\overline{N_1} N_1 \to W^+ W^-$ and hence the phenomena occurs upto a limiting point, which is $M_1\simeq 400$ GeV for $\sin\theta=0.2$ as shown in Fig. (\ref{fig:m1-m2}). Note that, Fig. (\ref{fig:m1-m2}) also shows the $Z$ and $H$ resonance at the left side, which is mostly independent of $M_2$ mass. However, they are highly disfavoured from $Z$ and $H$ invisible decays as they put even stronger constraint on $\sin\theta$.

\begin{figure}[thb]
$$
\includegraphics[height=5.5cm]{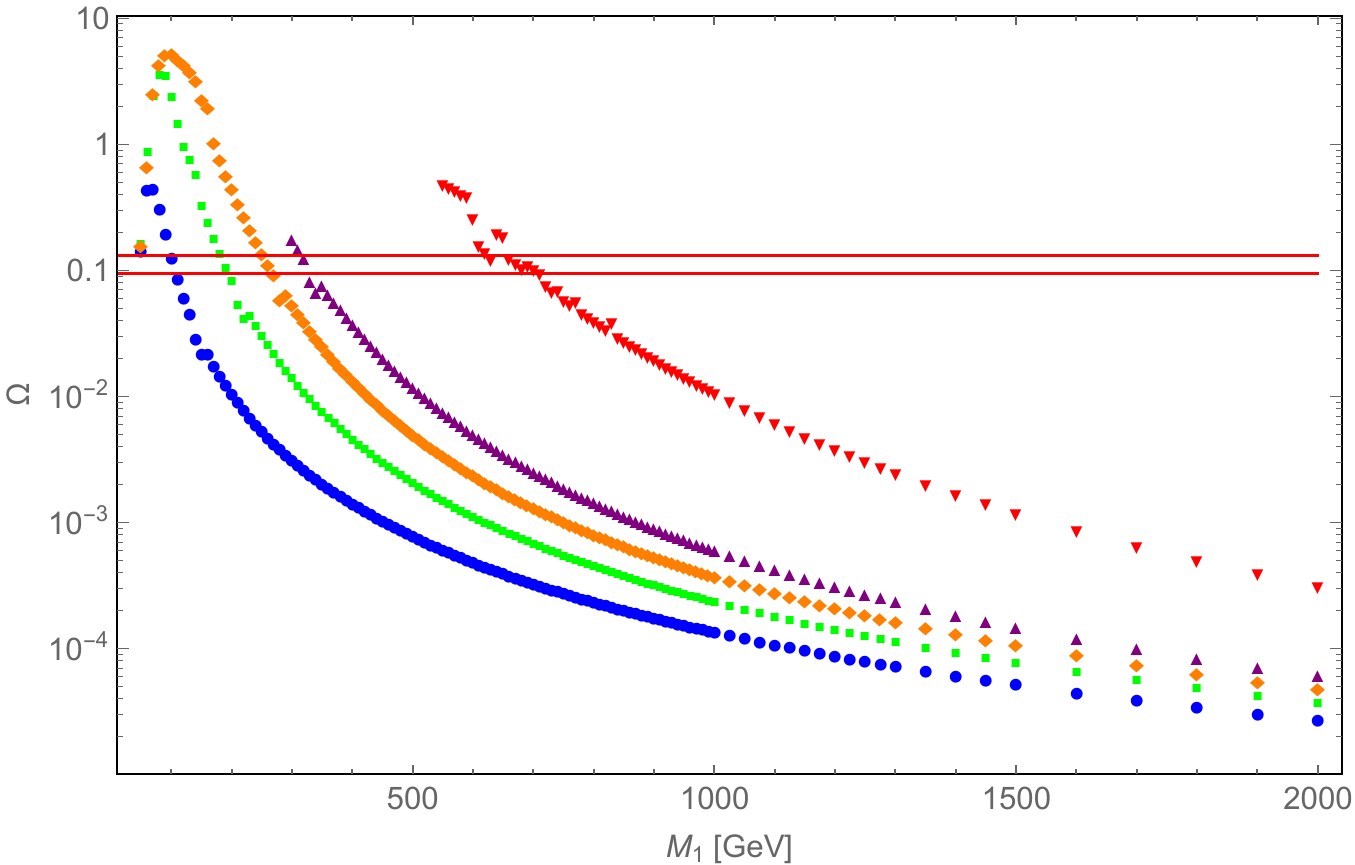}
\includegraphics[height=5.5cm]{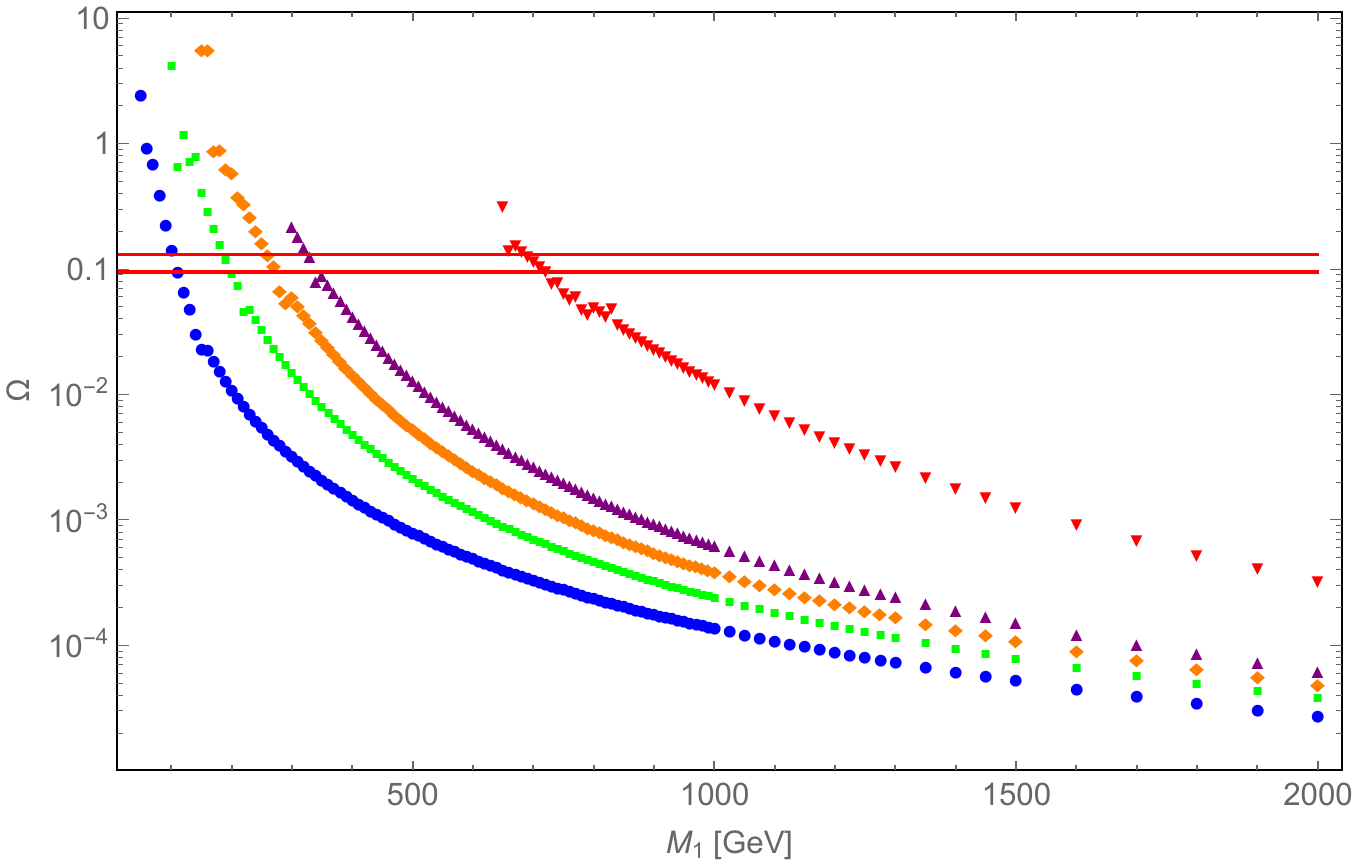}
$$
\caption{ Left: $\Omega$ versus Dark matter mass $M_1$ (in GeV) for $\sin\theta=0.1$ and $\Delta M = \{10,20,30,40,100\}${\rm GeV} (blue, green, orange, purple, red respectively). Right: Same as left but in terms of $\sin\theta=0.0001$. Horizontal lines indicate correct density.}
\label{fig:Omega-m3}
\end{figure}

In order to explore more on the coannihilation for smaller $\sin \theta$ region, we plot Fig. \ref{fig:Omega-m3}. One can see the $\Delta M$ dependency 
on relic density for a specific choice of mixing angle. In the left panel we choose $\sin \theta=0.1$ and that in the right $\sin \theta=0.0001$. The 
slices with constant $\Delta M$ is shown for $\Delta M=\{10,20,30,40,100\} {\rm GeV}$ in Blue, Green, Orange, Purple, Red respectively. We note here, 
that with larger $\Delta M$, the annihilation cross-section increases due to enhancement in Yukawa coupling ($Y \propto \Delta M$). However, co-annihilation 
decreases due to increase in $\Delta M$ as $\sigma \propto e^{-\Delta M}$. As co-annihilation dominates in this region of parameter space, the 
decrease in co-annihilation cross-section is much more than the increase in the annihilation channels, eventually leading to a larger relic density 
with increasing $\Delta M$ for a given value $M_1$. Hence bigger is the $\Delta M$, larger is the DM mass required to satisfy relic density with mixing 
angle $\sin \theta \le 0.1$. 

We choose a few benchmark points for further analysis of the model which characterizes different regions of the parameter space allowed by 
relic density and direct search. The details of those points are summarized in Table \ref{tab:bps}. One can see that for BP1 and BP3 points, 
coannihilation dominates through $N^{+} N^{-} \to W^{+} W^{-}$, followed by $N_2 \bar{N}_2 \to W^{+} W^{-}$ with significant other contributions 
coming from $~N^{-} ~\bar{N}_2 \to Z W^{-}$. Note that this is a generic feature for all such points satisfying relic density for small 
$Y=\Delta M \sin 2\theta/2v$ and small mixing angle, being independent of the dark matter mass. On the other hand, BP2 and BP4 represent 
`similar' points with larger Yukawa and mixing angle, where DM annihilation itself contributes fully to relic density. The significant 
contribution to relic density come from $N_1 \bar{N}_1 \to Z h, W^{+} W^{-}, hh$. However, for these points, the direct search cross-section 
is often large and in conflict with the LUX limit.

\begin{table}[htb]
\begin{center}
\begin{tabular}{|c|c|c|c|c|c|}
\hline
\hline

 Benchmark Points & BP1 & BP2 & BP3 & BP4 \\
 
\hline

\multirow{5}{*} {Parameters} &  $M_1= 150$,  & $M_1= 150$ & $M_1= 325$ & $M_1= 200$\\
                                              &  $M_2= 165$,  & $M_2= 650$, & $M_2= 365$, & $M_2=300$\\ 
                                              &  $M_N= 164.8$, & $M_N= 630$, & $M_N= 364.6$, & $M_N=293$\\
                                              &  $Y=0.006$,  &    $Y= 0.4$, &    $Y= 0.02$, &    $Y=0.1$\\   
                                              & $\sin\theta=0.1$, &   $\sin\theta=0.2$, &   $\sin\theta= 0.1$, &   $\sin\theta=0.26$\\ 
                                                 
\hline

$\Omega h^2$ & 0.096 & 0.093 & 0.11 & 0.104\\

\hline

\multirow{4}{*}{Processes} & $N^{+} N^{-} \to W^{+} W^{-}:53\%$ & $N_1 \bar{N}_1 \to Zh:36 \%$ & $N^{+} N^{-} \to W^{+} W^{-}:56\%$ & $N_1 \bar{N}_1 \to W^{+} W^{-}:81 \%$\\
                                                             & $N_2 \bar{N}_2 \to W^{+} W^{-}:11 \%$&  $N_1 \bar{N}_1 \to W^{+} W^{-}:30\%$& $N_2 \bar{N}_2 \to W^{+} W^{-}:11 \%$& $N_1 \bar{N}_1 \to Z h: 11\%$\\
                                                             & $N^{+} {N_2} \to Z W^{+}:10\%$& $N_1 \bar{N}_1 \to h h: 28\%$ & $N^{-} \bar{N_2} \to Z W^{-}:11\%$ & $N_1 \bar{N}_1 \to d \bar{d}: 1\%$\\
                                                             & $N^{+} {N_2}  \to \gamma W^{+}:6\%$& $N_1 \bar{N}_1 \to Z Z: 2\%$ &$ {N_1} \bar{N_2}  \to W^{+}W^{-}:5\%$ & $N_1 \bar{N}_1 \to b \bar{b}: 1\%$\\
\hline

$\sigma^{p}_{SI} (pb)$ & $4.1\times 10^{-9}$ & $1.52\times 10^{-8}$ & $4.26\times 10^{-9}$ & $2.04\times 10^{-7}$\\

\hline
\hline
\end {tabular}
\end{center}
\caption {Benchmark points of the dark matter model. Parameters at this benchmark point, relic density, relative contributions from annihilation channels and spin independent direct detection cross-section with proton (in pb) has been mentioned.}
\label{tab:bps}
\end{table}

\section{Electroweak Precision tests and direct limits on vector-like leptons}
Any vector like fermion doublet beyond SM framework contribute to the electroweak precision test parameters $S$, $T$ and $U$~\cite{peskin,EW_constraint1}. 
In fact, a more generalized set of parameters for electroweak precision test are $\hat{S}$, $\hat{T}$, $W$ and $Y$~\cite{EW_constraint2}, where 
the $\hat{S}$, $\hat{T}$ are related to Peskin-Takeuchi parameters $S$, $T$ as $\hat{S}=\alpha S/4 \sin^2 \theta_w$, $\hat{T}=\alpha T$, 
while $W$ and $Y$ are two new set of parameters. The observed values of these parameters at LEP-I and LEP-II set a lower bound on the 
mass scale of new fermions. Global fit of the electroweak precision parameters for a light Higgs~\cite{EW_constraint2}~\footnote{The value 
$\hat{S}$, $\hat{T}$, $W$ and $Y$ are obtained using a Higgs mass $m_h=115$ GeV. However, we now know that the SM Higgs mass is 125 GeV. Therefore, 
the value of $\hat{S}$, $\hat{T}$, $W$ and $Y$ are expected to change. But this effect is nullified by the small values of $\sin \theta$.} is shown 
in the following Table.
\begin{table}[h]
\begin{center}
\begin{tabular}{|c|c|c|c|c|}
\hline 
  & $10^3 \hat{S}$ & $10^3 \hat{T}$ & $10^3 W$ & $10^3 Y$ \\\hline
Light Higgs & $0.0\pm 1.3$ & $0.1 \pm 0.9$ & $0.1 \pm 1.2$ & $-0.4\pm 0.8$ \\ \hline
\end{tabular}
\end{center}
\caption{Global fit for the electroweak precision parameters taken from ref.~\cite{EW_constraint2}.}
\end{table} 
In the present case, we have two neutral fermions $N_1$ and $N_2$ and a charged fermion $N^-$. Note that $N_1$ is dominantly a singlet and 
a small admixture of doublet component, while $N_2$ is dominantly a doublet and a small admixture of singlet component. Therefore, the 
mixing is important for their contribution to $\hat{S}$, $\hat{T}$, $W$ and $Y$. In terms of $M_1$, $M_2$, $M^\pm$ and $\sin \theta$ we can 
compute $\hat{S}$ as~\cite{Cynolter:2008ea}:
\begin{eqnarray}
\hat{S} &=& \frac{g^2}{16\pi^2} \left[ \frac{1}{3} \left\{  \ln \left(\frac{\mu^2}{({M^{\pm}})^2} \right) -\cos^4 \theta \ln \left(\frac{\mu^2}{M_2^2} \right)
- \sin^4 \theta \ln \left(\frac{\mu^2}{M_1^2} \right) \right\} - 2\sin^2 \theta \cos^2 \theta \left\{   
\ln \left(\frac{\mu^2}{M_1 M_2} \right) \right.\right.\nonumber\\ 
 && \left. \left. + \frac{M_1^4-8 M_1^2 M_2^2 +M_2^4}{9 (M_1^2-M_2^2)^2} + \frac{(M_1^2+M_2^2)(M_1^4-4M_1^2M_2^2+M_2^4)}{6(M_1^2-M_2^2)^3} 
\ln \left(\frac{M_2^2}{M_1^2} \right) \right.\right.\nonumber\\
&& \left. \left. +   \frac{M_1 M_2(M_1^2 + M_2^2)}{2(M_1^2-M_2^2)^2} + \frac{M_1^3M_2^3}{(M_1^2-M_2^2)^3} \ln \left(\frac{M_2^2}{M_1^2} \right)      \right\}          \right]
\end{eqnarray}
where $\mu$ is at the EW scale. 
\begin{figure}[htbp]
\includegraphics[scale=.3]{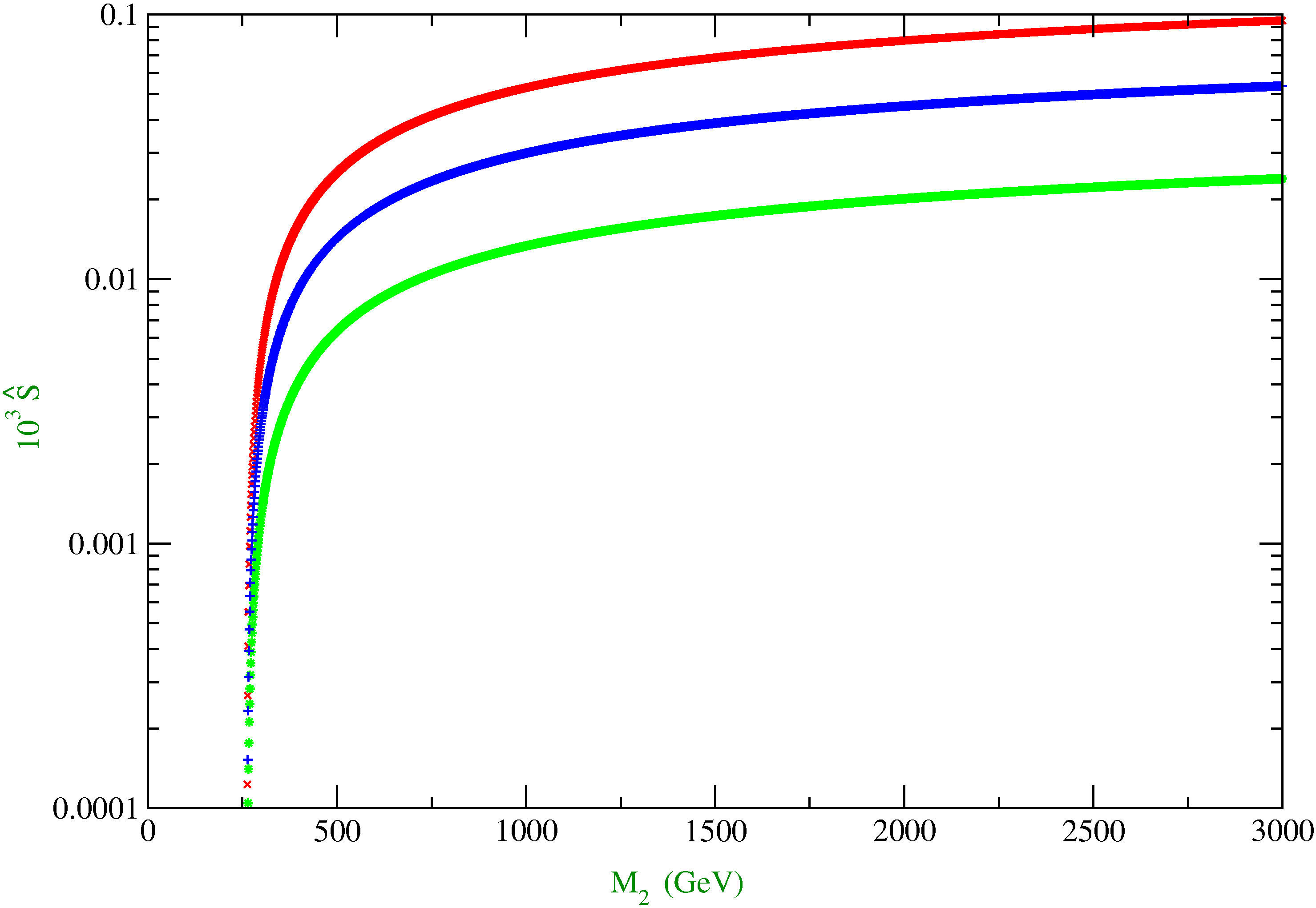}
\includegraphics[scale=0.3]{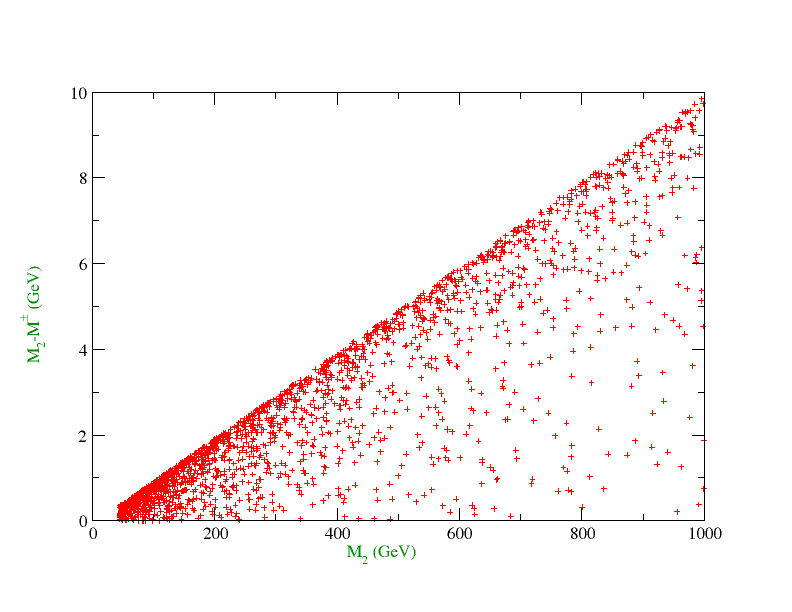}
\caption{In the left panel, $\hat{S}$ is shown as a function of $M_2$ for $M_1=150$ GeV and $\rm sin\theta = 0.05$ (Green colour, bottom), 
$\rm sin\theta = 0.075$ (Blue color, middle) and $\rm sin\theta = 0.1$ (Red color, top). In the right panel, allowed values of $\hat{S}$ in the plane 
of $M_2-M^\pm$ versus $M_2$ for $\sin \theta=0.05$.}\label{shat_values}
\end{figure}
In left-panel of Fig. (\ref{shat_values}), we have shown $\hat{S}$ as a function of $M_2$ for different values of the mixing angles while keeping $M_1=150$ GeV. 
On the other hand, in the right panel, we have shown the allowed values of $\hat{S}$ in the plane of $M_2-M^\pm$ versus $M_2$ for $\sin \theta=0.05$. We observed 
that $\hat{S}$ does not put strong constraints on $M_1$ and $M_2$. Moreover, small values of $\sin \theta$ allows a small mass splitting between $N_2$ and $N^-$ 
which relaxes the constraint on $\hat{T}$ parameter as we discuss below. In terms of $M_1$, $M_2$, $M^\pm$ and $\sin \theta$ one can compute $\hat{T}$ as~\cite{Cynolter:2008ea}:  
\begin{equation}
\hat{T}=\frac{g^2}{16 \pi^2 M_W^2}\left[2 \sin^2 \theta \cos^2 \theta ~\Pi(M_1,M_2,0)-2\cos^2 \theta ~\Pi(M^\pm,M_2,0)-2\sin^2\theta ~\Pi(M^\pm,M_1,0)\right]\,,
\end{equation}
 where $\Pi(a,b,0)$ is given by: 
\begin{eqnarray}
\Pi (a,b,0) &=& -\frac{1}{2}(M_a^2+M_b^2)\left({\rm Div}+\ln \left(\frac{\mu^2}{M_a M_b} \right)  \right)-\frac{1}{4} (M_a^2+ M_b^2)-\frac{(M_a^4+M_b^4)}
{4(M_a^2-M_b^2)} \ln \frac{M_b^2}{M_a^2} \nonumber\\
&& + M_a M_b \left\{{\rm Div}+ \ln  \left(\frac{\mu^2}{M_a M_b} \right)+1+ \frac{(M_a^2 +M_b^2)}{2(M_a^2-M_b^2)} \ln  \frac{M_b^2}{M_a^2}  \right\}
\end{eqnarray}
\begin{figure}[htbp]
\includegraphics[scale=.3]{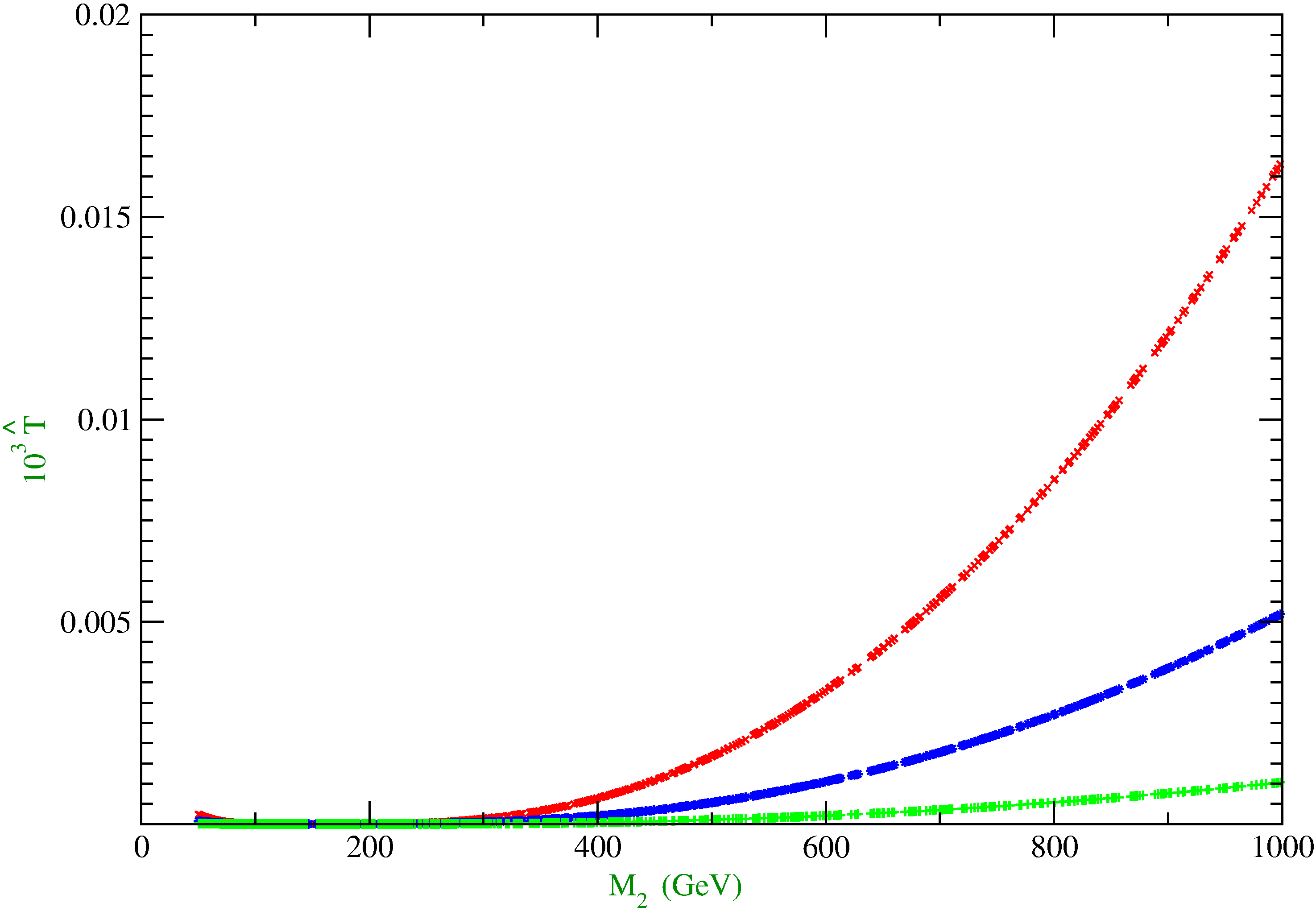}
\includegraphics[scale=0.3]{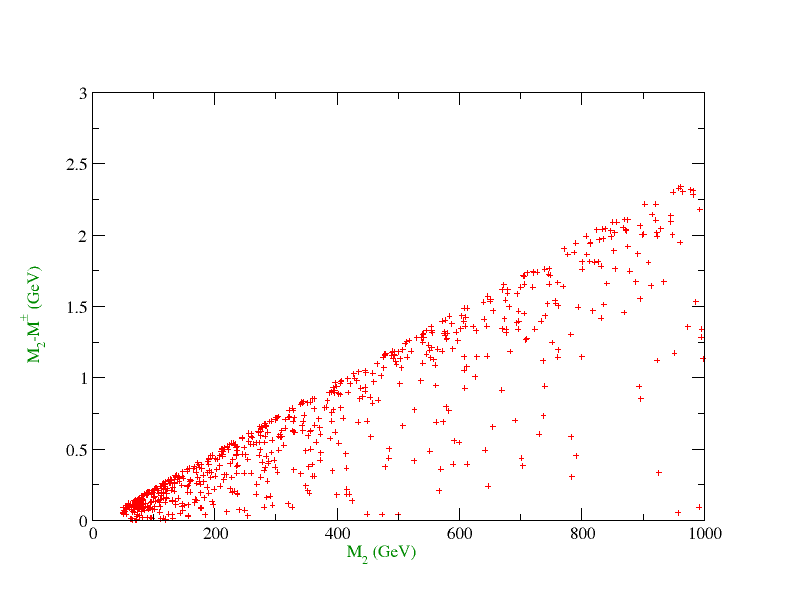}
\caption{In the left panel, $\hat{T}$ is shown as a function of $M_2$ for $M_1=150$ GeV and $\rm sin\theta = 0.05$ (Green colour, bottom), 
$\rm sin\theta = 0.075$ (Blue color, middle) and $\rm sin\theta = 0.1$ (Red color, top). In the right panel, allowed values of $\hat{T}$ in the plane 
of $M_2-M^\pm$ versus $M_2$ for $\sin \theta=0.05$.}\label{that_values}
\end{figure}
From the left panel of Fig. (\ref{that_values}) we see that for $\sin \theta < 0.05$ we don't get strong constraints on $M_2$ and $M_1$. Moreover, 
small values of $\sin \theta$ restricts the value of $M_2-M^\pm$ to be less than a GeV. As a result large $M_2$ values are also allowed. Near $M_2 
\approx M^\pm$, $\hat{T}$ vanishes as expected. The value of $Y$ and $W$ are usually suppressed by the masses new fermions. Since the allowed masses 
of $N_1$, $N_2$ and $N^\pm$ are above 100 GeV by the relic density constraint, so $Y$ and $W$ are naturally suppressed.

\section{Collider signature}
 
If the new leptons are $\simeq 500$ GeV, they can be produced at the Large Hadron Collider (LHC). They will eventually decay to the lightest stable particle $N_1$. The DM $N_1$ is stable, charge neutral and will escape from the detector, while its charged partner $N^\pm$ may give promising signature if it is produced. For example, 
$N^\pm$ can be pair produced via the Drell-Yan process mediated by $\gamma$ and $Z$-boson. Note that the production of $N^\pm$ is independent 
of singlet-doublet mixing. So the small values of $\sin \theta$, required for evading Xenon-100 and LUX bound at direct detection of DM, does 
not affect the pair production of $N^\pm$. On the other hand, production of $N_1 N^\pm$ pair via the exchange of SM $W^\pm$ will be suppressed 
by low values of $\sin\theta$. Therefore, in what follows we will discuss signature of vector-like charged fermions $N^\pm$, pair produced mainly 
through $\gamma$ and $Z$ mediated Drell-Yan process. 
\begin{figure}[htbp]
\includegraphics[height=5.5cm]{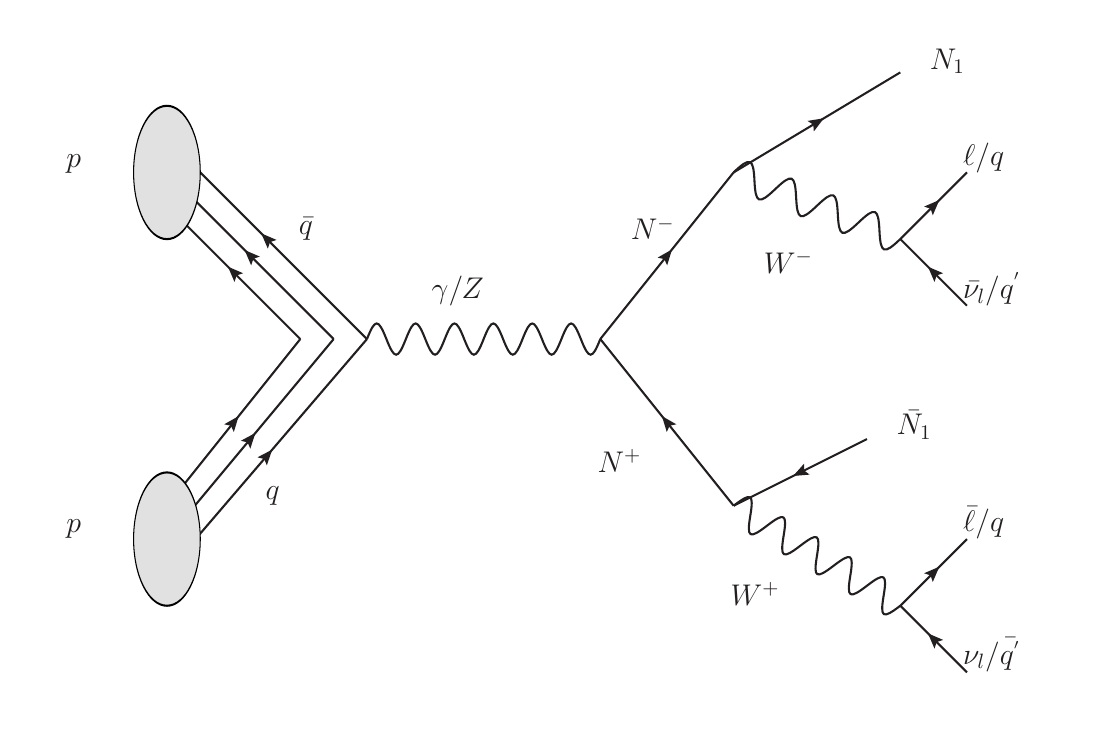}
\includegraphics[height=5.5cm]{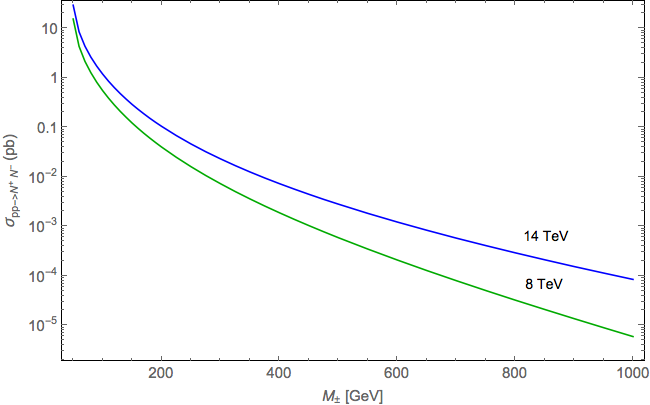}
\caption{Left panel: Feynman graph producing $N^+N^-$ pair at LHC and its subsequent decays. Right panel: Variation in production cross section $\sigma_{pp\to N^+ N^-}$ (pb) at LHC with respect to $M_{\pm}$ for $E_{cm}= 8$ TeV (Green, below) 
and $E_{cm}= 14$ TeV (Blue, above).}
\label{fig:FD-collider}
\end{figure}

Once the $N^\pm$ is produced it decays via $N^\pm \to N_1 W^\pm$. If the mass splitting between $N^\pm$ and $N_1$, which is equivalent to 
$\Delta M= M_2 - M_1 \equiv M^\pm - M_1$, is larger than $W$ mass, then the two body is favorable, otherwise the decay will proceed through off-shell $W$. So the relevant signatures in case of pair production of $N^+N^-$ at LHC will be as follows: $pp\to N^+ N^-\to \overline{N_1} N_1 W^+ W^-$; subsequently the possible final states are:  
\begin{enumerate}
\item One lepton + Di-jet + Missing energy  ($\ell 2jE_{T}\!\!\!\!/$)
\item Two oppositely charged leptons + Missing energy ($2\ell E_{T}\!\!\!\!/$)
\item 4 jets + Missing energy ($4jE_{T}\!\!\!\!/$)
\end{enumerate}
depending on whether the W's decay hadronically or leptonically. See for example, the Feynman graph in the left of Fig. (\ref{fig:FD-collider}). We also show the variation in production cross-section $\sigma_{pp\to N^+ N^-}$ (pb) for $N^+N^-$ at LHC with respect to $M_{\pm} (GeV)$ for $E_{cm}= 8$ TeV (Green, below) and $E_{cm}= 14$ TeV (Blue, above). Accordingly, we tabulate in Table (\ref{tab:signal-cs}) the production cross-sections as well as the cross-sections in hadronically quiet dilepton final state as well as in single lepton state for the benchmark points chosen above. For reference, contributions to the leptonic final states from dominant SM background; namely 
$W^+W^-, ZZ, t\bar{t} $ has also been tabulated in Table (\ref{tab:back-8}) and Table (\ref{tab:back-14}) for $E_{cm}= 8$ TeV and $E_{cm}= 14$ TeV respectively. It is seen that the only way to tame down the background is to put a very high missing energy cut, $E_{T}\!\!\!\!/>100$ GeV. The amount of signal that will be left after MET cut, depends on the amount of transverse momentum that is transferred to $N_1$ from the decay of $N^{\pm}$, which is proportional to the mass difference. However, as $W^{\pm}$ is less massive than the DM, the significant part of the momentum will be carried by the DM. hence, it is expected that the peak of MET will be much higher than 100 GeV and a cut hence will retain a significant part of the signal. CalcHep \cite{calchep} and 
Pythia \cite{pythia} event generators have been used to produce the cross-sections. Just for clarifications, we also note here that missing energy is identified in terms 
of the visible momenta as follows: vector sum of the x and y components of the momenta separately for all visible objects form visible transverse 
momentum $(p_T)_{vis}$ and that is precisely the missing energy from momentum conservation.
\bea
(p_T)_{vis}=\sqrt{(\sum p_x)^2+(\sum p_y)^2}=E_{T}\!\!\!\!/ 
\eea
where, $\sum p_x =\sum (p_x)_{\ell}+\sum (p_x)_{jet}$ and similarly for $\sum p_y$.
 
 What we see from the table, is that BP1 has a very small $M_N$ and hence results with a huge cross-section. Hence, this point lies close to what has been discarded from non-observation of any excess in semi-leptonic or leptonic channels so far from 7 TeV data at LHC. However, for BP3 and BP4, there is a strong possibility that one might see an excess in the next run of LHC after careful background reduction. While for BP2, it seems very hard to see any excess in above channels unless we go to high luminosity LHC. The analysis presented here is more indicative than exhaustive.  For generic collider implications of vector like dark matter, see \cite{vdm-recent}, which also imply additional constraints on the DM parameter space.

\begin{table}[htb]
\begin{center}
\begin{tabular}{|c|c|c|c|c|c|c|}
\hline
\hline

 Benchmark Points & ${[\sigma_{pp\to N^+ N^-}]}_8$  & ${[\sigma_{\ell 2jE_{T}\!\!\!\!/}]}_8$ & 
 ${[\sigma_{2\ell E_{T}\!\!\!\!/}]}_8$ & ${[\sigma_{pp\to N^+ N^-}]}_{14}$ & ${[\sigma_{\ell 2jE_{T}\!\!\!\!/}]}_{14}$ & ${[\sigma_{2\ell E_{T}\!\!\!\!/}]}_{14}$\\
 
\hline

BP1 & 284 & 80 & 12.5 & 700 & 197 & 31 \\
                                                 
\hline

BP2 & 0.58 & 0.16 & 0.025 & 3.3 & 0.93 & 0.15 \\

\hline

BP3 & 10.13 & 2.85 & 0.45 & 35.1 & 9.88 & 1.55 \\

\hline 

BP4 & 27.02 & 7.6 & 1.19 & 82.5 & 23.2 & 3.64 \\

\hline
\hline
\end {tabular}
\end{center}
\caption {Production Cross-sections $\sigma_{pp\to N^+ N^-}$ for the benchmark points at LHC for $E_{cm}= 8$ and 14 TeV. The leptonic final states $\sigma_{\ell 2jE_{T}\!\!\!\!/}$ and $\sigma_{2\ell E_{T}\!\!\!\!/}$ are also mentioned. All cross-sections are in fb.}
\label{tab:signal-cs}
\end{table}

\begin{table}[htb]
\begin{center}
\begin{tabular}{|c|c|c|c|c|}

\hline
\hline

SM Background & ${[\sigma_{\ell 2jE_{T}\!\!\!\!/}]}_8$ & ${[\sigma_{\ell 2jE_{T}\!\!\!\!/}]}_8, E_{T}\!\!\!\!/>100$ & ${[\sigma_{2\ell E_{T}\!\!\!\!/}]}_8$ & ${[\sigma_{2\ell E_{T}\!\!\!\!/}]}_8, E_{T}\!\!\!\!/>100$ \\  
\hline

WW & 2.30 & 1.04 & 0.74 & $\le 0.35$  \\
                                                 
\hline

ZZ & 0.38 & 0.005 & 10.4 & $\le 0.5$  \\

\hline

$t\bar{t}$ & 5.43 & 0.16 & 0.09 & $\simeq 0$\\

\hline
\hline
\end {tabular}
\end{center}
\caption { SM background at LHC for $E_{cm}= 8$ TeV for ${\ell 2jE_{T}\!\!\!\!/}$ and ${2\ell E_{T}\!\!\!\!/}$ channels before and after 
missing energy cut $E_{T}\!\!\!\!/>100$ GeV. All cross-sections are in fb.}
\label{tab:back-8}
\end{table}

\begin{table}[htb]
\begin{center}
\begin{tabular}{|c|c|c|c|c|}

\hline
\hline

SM Background & ${[\sigma_{\ell 2jE_{T}\!\!\!\!/}]}_{14}$ & ${[\sigma_{\ell 2jE_{T}\!\!\!\!/}]}_{14}, E_{T}\!\!\!\!/>100$ & ${[\sigma_{2\ell E_{T}\!\!\!\!/}]}_{14}$ & ${[\sigma_{2\ell E_{T}\!\!\!\!/}]}_{14}, E_{T}\!\!\!\!/>100$ \\  
\hline

WW & 4.37 & 0.027 & 1.26 & $\le 0.7$  \\
                                                 
\hline

ZZ & 0.83 & 0.01 & 16.9 & $\le 1$  \\

\hline

$t\bar{t}$ & 21.4 & 0.88 & 0.41 & $\le 0.1$\\

\hline
\hline
\end {tabular}
\end{center}
\caption { SM background at LHC for $E_{cm}= 14$ TeV for ${\ell 2jE_{T}\!\!\!\!/}$ and ${2\ell E_{T}\!\!\!\!/}$ channels before and after missing energy cut $E_{T}\!\!\!\!/>100$ GeV. All cross-sections are in fb.}
\label{tab:back-14}
\end{table}

\begin{figure}[htbp]
\includegraphics[height=6.5cm]{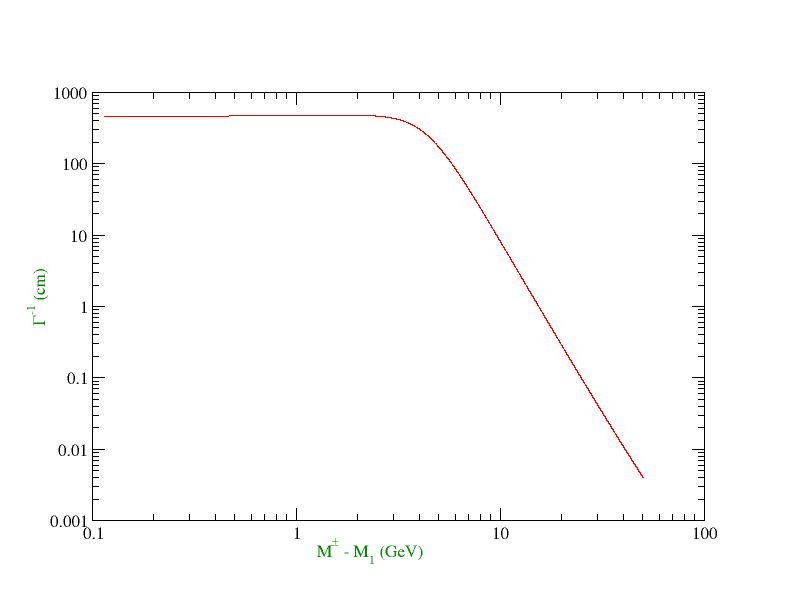}
\includegraphics[height=6.5cm]{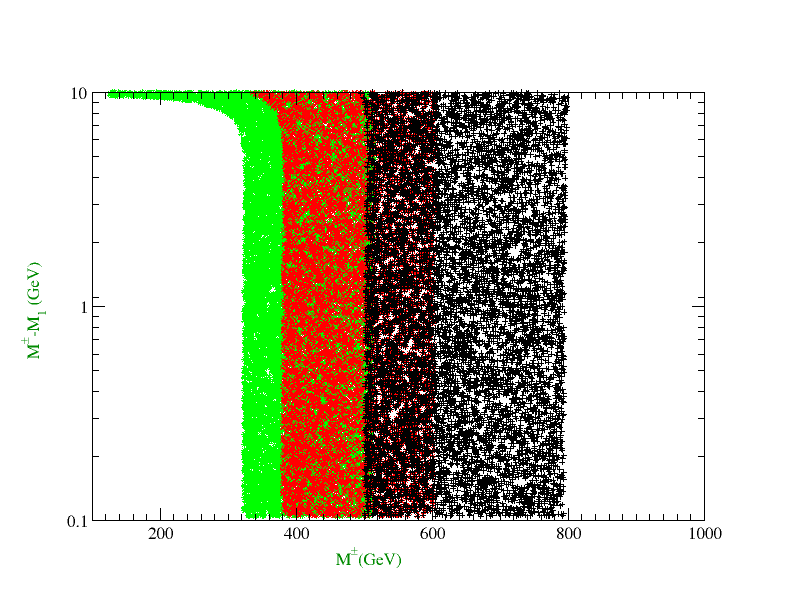}
\caption{Left panel: Displaced vertex of $N^-$ for $M^\pm=150$ GeV, $m_\ell=105$ MeV and $\sin \theta=3 \times 10^{-4}$. 
Right panel: $\Gamma^{-1}$ values varying between (1 - 10) cm in the plane of $\Delta M$ versus $M^\pm$ for $\sin \theta= 3\times 10^{-4}$ 
(in Green), $2\times 10^{-4}$ (in Red) and $10^{-4}$ (in Black) simultaneously from left to right.}\label{displaced_vertex}
\end{figure}

There is another very interesting signature of the model. For example, if the mass splitting between $N^\pm$ and $N_1$ is less than 90 GeV, then $N^-$ will decay via three body suppressed process: $N^- \to N_1 \ell \nu_\ell$ and 
$N^- \to N_1 + {\rm di-jets}$, due to small values of $\sin \theta$. The latter one may not be a suitable process to search at LHC, while 
the former one is useful to look for via a displaced vertex signature as discussed below. The decay rate can be given as:
\begin{equation}
\Gamma = \frac{ G_F^2 sin^2\theta}{24 \pi^3} M_N^5  I
\end{equation}
where $G_F$ is the Fermi coupling constant and 
\begin{equation}\label{decay-rate}
I=\frac{1}{4}\lambda^{1/2}(1,a^2,b^2) F_1(a,b) + 6 F_2 (a,b)\ln  \left(\frac{2a}{1+a^2-b^2-\lambda^{1/2}(1,a^2,b^2)} \right) \,. 
\end{equation}
In the above Equation $F_1 (a,b)$ and $F_2 (a,b)$ are two polynomials of $a=M_1/M_N$ and $b=m_\ell/M_N$, where $m_\ell$ is the 
charged lepton mass. Up to ${\cal O}(b^2)$, these two polynomials are given by
\begin{eqnarray}
F_1 (a,b) &=& \left( a^6-2a^5-7a^4(1+b^2)+10a^3(b^2-2)+a^2(12b^2-7)+(3b^2-1)\right)\nonumber\\
F_2 (a,b) &=&  \left(a^5+a^4+a^3(1-2b^2)\right)\,.
\end{eqnarray} 
In Eq. (\ref{decay-rate}), $\lambda^{1/2}=\sqrt{1+a^4+b^4-2a^2-2b^2-2a^2b^2}$ defines the phase space. In the limit $b=m_\ell/M_N 
\to 1-a=\Delta M/M_N$, $\lambda^{1/2}$ goes to zero and hence $I\to 0$. In the left-panel of Fig. (\ref{displaced_vertex}), we have shown $\Gamma^{-1}({\rm cm})$ as a function of $\Delta M$ by taking $M_N=150$ GeV and $m_\ell=150$ MeV. We see that for small 
$\Delta M$, say $\Delta M < 10$ GeV, we get a displaced vertex more than 1 cm. In the right panel of Fig.(\ref{displaced_vertex}), 
we show $\Gamma^{-1}$ values varying between (1 - 10) cm in the plane of $\Delta M$ versus $M_N$ for $\sin \theta= 3\times 10^{-4}$ 
(in Green), $2\times 10^{-4}$ (in Red) and $10^{-4}$ (in Black) simultaneously from left to right. The important point to be noted here is that to get a large displaced vertex we need a small mixing angle between the singlet and doublet. In fact, the small mixing angle is favored by all the constraints we discussed in previous sections. More over, the small mixing angle also does not hamper the relic abundance of dark matter as summarised in the right panel of Fig. (\ref{fig:Omega-m3}).

\section{Conclusions and outlook} 

\begin{figure}[thb]
$$
\includegraphics[height=8.0cm]{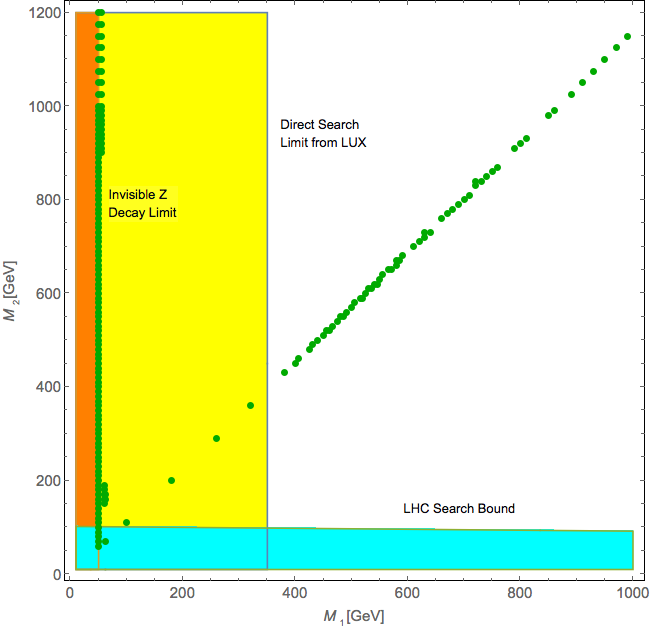}
$$
\caption{Summary of all constraints in $M_1-M_2$ (GeV) parameter space with $\sin \theta=0.1$ from relic density (green dots), direct search (yellow region is forbidden by LUX), invisible Z-decay (Orange region is forbidden) and collider (LHC) search limit (cyan region is disallowed).}
\label{fig:summary}
\end{figure}

We discussed the possibility of a minimal vector-like fermion DM in a simple beyond the Standard Model framework. The SM is extended by two vector-like 
leptons $N$, an $SU(2)$ doublet, and $\chi_0$, a singlet, both odd under $Z_2$ resulting to three physical states $N_1=\cos \theta \chi^0 + 
\sin \theta N^0$ and $N_2=\cos \theta N^0 - \sin \theta \chi^0$ and $N^{\pm}$ the charged partner of $N^0$, where $\theta$ is the mixing angle. 
 $N_1$ is a suitable DM candidate and the relic density has been evaluated. Relic density and direct search of this DM is crucially guided by the interplay of Yukawa ($Y$) and $SU(2)$ gauge interactions of the DM which are functions of ${M_1,M_2,\sin\theta}$. It turns out that it satisfies WMAP/PLANCK constraints in a significantly 
large region of parameter space for its mass $M_1 \gsim 45 $ GeV. The relevant constraints on $\sin \theta$ from the invisible $Z$ and Higgs decay, electroweak precision data and direct detection of dark matter are discussed. For $M_1 < 45$ GeV, $N_1$ is strongly constrained by the invisible Z-decay width, while for $M_1 > 45 $ GeV, the direct detection of $N_1$ DM at Xenon-100 and LUX give the strongest constraint on $\sin \theta$, thus ruling out its viability for $\sin \theta \gsim  0.1$. However, correct relic density of DM can also be found for larger mixing ($\sin \theta$) and appropriate $\Delta M$ to adjust the Yukawa interaction $Y=(\Delta M \sin\theta \cos \theta)/v$ with the $SU(2)$ interaction of the DM. As an example, we can assimilate all the constraints together for a specific choice of $\sin \theta$ into the plane of $M_1-M_2$. This is what we have shown in Fig. \ref{fig:summary} by considering:

\begin{align*}
\label{eq:constraints}
\nonumber
\rm{Invisible ~Z~ decay}: M_1< \frac{M_z}{2}\sim 45 ~GeV & \to \sin \theta \lesssim 0.00125 \\
\nonumber
\rm{Relic ~Density}: M_2 \lesssim M_1+100~ GeV, & ~for~ \sin \theta \lesssim 0.1\\
\nonumber
\rm{Direct ~Search}: M_1 \ge 350 ~GeV, & ~for~ \sin \theta \sim 0.1\\
\rm{Collider~ Bound}: M_2 \simeq M^{\pm} \ge 101 ~GeV & ~for~ \sin \theta \sim 0.1
\end{align*}
 We plot it for $\sin\theta=0.1$, as it does satisfy most of the constraints discussed here. Note that for $\sin \theta \lesssim 0.1$, 
constraints from invisible Higgs decay is almost negligible. From Fig. (\ref{fig:summary}), We see that a large part of the parameter space is 
still allowed except for the yellow band (disfavoured by direct search), orange band (disfavoured by the Invisible $Z$ decay) and cyan band 
(disfavoured by direct collider search \cite{collider}).

The DM is elusive at the collider and simply escapes without detection. However, we found that the charged companion of the DM, {\it i.e.} $N^\pm$ can 
give interesting signature including opposite sign dileptons associated with jets or hadronically quiet. In particular, we showed that the three body 
decay of $N^\pm$ can give large displaced vertex signature at LHC provided that the mass splitting between $N_1$ and $N^\pm$ is less than around 10 GeV. 
But the observable displaced vertex of $N_1$ needs $\sin \theta \lsim 0.01$ for which the spin independent DM-nucleon cross-section is much less than 
the sensitivity of Xenon1T. Thus, the signature of DM at direct detection experiments seems to be complementary to its collider signature in terms of 
displaced vertex. However, the individual signatures seems to be appealing. 

The detection of DM via indirect search is also possible, though a full discussion was beyond the scope of this draft. For example, the annihilation 
of $\overline{N_1} N_1 \to h h$ and $\overline{N_1} N_1 \to h Z$ and subsequently $h \to \gamma \gamma $ can lead to gamma ray excess in the galactic 
center. In other words, Fermi-Lat data from galactic center can be used to constrain the Higgs production from the DM annihilation in our 
model~\cite{Bernal:2012cd}. We will discuss this issue in a future publication.    

To summarise, the model presented here is a minimal one for a renormalisable vector-like leptonic DM that lies within the scope of detection. Neither 
the neutral component of the doublet nor the neutral singlet can individually qualify for a renormalisable vector-like leptonic DM, while their combination 
does.

\section{Acknowledgements}

SB would like to acknowledge to the DST-INSPIRE grant no PHY/P/SUB/01 at IIT Guwahati and to the local hospitality provided at IIT- Hyderabad during 
a visit where the major part of the work was carried. NS is partially supported by the Department of Science and Technology, Govt. of India under
the financial Grant SR/FTP/PS-209/2011.


\end{document}